\documentclass[final]{jfm}

\usepackage{graphicx}
\usepackage{natbib}
\usepackage{MR_JFM}
\usepackage[latin1]{inputenc}
\usepackage{color}
\usepackage{mathtools}

\ifCUPmtlplainloaded \else
  \checkfont{eurm10}
  \iffontfound
    \IfFileExists{upmath.sty}
      {\typeout{^^JFound AMS Euler Roman fonts on the system,
                   using the 'upmath' package.^^J}%
       \usepackage{upmath}}
      {\typeout{^^JFound AMS Euler Roman fonts on the system, but you
                   dont seem to have the}%
       \typeout{'upmath' package installed. JFM.cls can take advantage
                 of these fonts,^^Jif you use 'upmath' package.^^J}%
      }
  \else
  \fi
\fi

\ifCUPmtlplainloaded \else
  \checkfont{msam10}
  \iffontfound
    \IfFileExists{amssymb.sty}
      {\typeout{^^JFound AMS Symbol fonts on the system, using the
                'amssymb' package.^^J}%
       \usepackage{amssymb}%
         \let\leq=\leqslant
         \let\geq=\geqslant
      }{}
  \fi
\fi

\ifCUPmtlplainloaded \else
  \IfFileExists{amsbsy.sty}
    {\typeout{^^JFound the 'amsbsy' package on the system, using it.^^J}%
     \usepackage{amsbsy}}
    {}
\fi

\newsavebox{\astrutbox}
\sbox{\astrutbox}{\rule[-5pt]{0pt}{20pt}}

\newcommand{\vect}[1]{{\bf {#1}}}
 
\newcommand\mnras{Monthly Notices of the Royal Astronomical Society} 
\newcommand\apj{The Astrophysical Journal} 

\title[Inertial waves in a differentially rotating spherical shell]{Inertial waves in a
differentially rotating spherical shell}

\author[C. Baruteau and M. Rieutord]%
{C.\ns B\ls A\ls R\ls U\ls T\ls E\ls A\ls U$^1$%
  \thanks{Email address for correspondence: C.Baruteau@damtp.cam.ac.uk}
  \and M.\ns R\ls I\ls E\ls U\ls T\ls O\ls R\ls D$^2$}

\affiliation{$^1$Department of Applied Mathematics and Theoretical Physics, University of Cambridge, Wilberforce Road, Cambridge CB3 0WA, United Kingdom\\[\affilskip]
$^2$Institut de Recherche en Astrophysique et Plan{\'e}tologie, CNRS et Universit{\'e} de Toulouse, 14 avenue E. Belin, 31400 Toulouse, France}

\begin{document}

\maketitle

\begin{abstract} 
  We investigate the properties of small-amplitude inertial waves
  propagating in a differentially rotating incompressible fluid
  contained in a spherical shell. For cylindrical and shellular
  rotation profiles and in the inviscid limit, inertial waves obey a
  second-order partial differential equation of mixed type. Two kinds
  of inertial modes therefore exist, depending on whether the
  hyperbolic domain where characteristics propagate covers the whole
  shell or not. The occurrence of these two kinds of inertial modes is
  examined, and we show that the range of frequencies at which
  inertial waves may propagate is broader than with solid-body
  rotation. Using high-resolution calculations based on a spectral
  method, we show that, as with solid-body rotation, singular modes
  with thin shear layers following short-period attractors still exist
  with differential rotation. They exist even in the case of a full
  sphere.  In the limit of vanishing viscosities, the width of the
  shear layers seems to weakly depend on the global background shear,
  showing a scaling in E$^{1/3}$ with the Ekman number E, as in the
  solid-body rotation case. There also exist modes with thin detached
  layers of width scaling with E$^{1/2}$ as Ekman boundary layers.
  The behavior of inertial waves with a corotation resonance within
  the shell is also considered. For cylindrical rotation, waves get
  dramatically absorbed at corotation. In contrast, for shellular
  rotation, waves may cross a critical layer without visible
  absorption, and such modes can be unstable for small enough Ekman
  numbers.
\end{abstract}


\section{Introduction}

The fluid flows that pervade the interior of stars and planets are
most of the time dominated by a strong background rotation. The
Earth's atmosphere, oceans, or its liquid core are typical
examples. Such flows are studied in a rotating frame, and the
dominance of rotation arises through that of the Coriolis acceleration
over all other terms in the momentum equation (the centrifugal
acceleration is usually combined with the pressure gradient through
the so-called reduced pressure). The Coriolis acceleration implies
that the response of a rotating fluid to small amplitude oscillations
may become singular.

It is known indeed since the work of \cite{cartan22} that the pressure
perturbations of an inviscid incompressible fluid obey Poincar{\'e}
equation, which has the remarkable property of being spatially
hyperbolic. Since boundary conditions have to be met by solutions,
oscillations of bounded rotating fluids are a mathematically ill-posed
problem. It means that, in most general containers, eigenmodes of
inviscid incompressible fluids may simply not exist. Mathematically,
the point spectrum of the operator is said to be empty. However, a few
containers such as the full ellipsoid or the cylindrical layer allow
the existence of such eigenmodes, because of the separation of spatial
variables that is possible in these geometries. Solutions were worked
out by \cite{Kelvin1880} for the cylinder, and by \cite{Bryan1888} for
the sphere and the ellipsoid. The singular nature of the solutions to
Poincar{\'e} equation have been uncovered almost a century later by
the works of \cite{StRic69} and \cite{stewar71,stewar72a}.
\cite{stewar72b} showed that these singularities should turn into
oscillating shear layers through viscosity.

New interest into the singular solutions to the Poincar{\'e} equation
was put forward with the work of \cite{ML95} on gravity modes in
two-dimensional basins. The pressure fluctuations of internal gravity
modes also satisfy Poincar{\'e} equation, and \cite{ML95} showed the
crucial role of characteristic trajectories and coined the concept of
attractor to describe the convergence of characteristics to special
limit cycles. At the same time, \cite{HK95} and \cite{K95}
investigated the internal shear layers spawned by the critical
latitude singularity (another singularity different from that of the
attractors), in relation with the precession flows of a spherical
shell.

Subsequent work by \cite{RV97}, \cite{RGV01} and \cite{RVG02}
demonstrated how attractors feature the oscillations of slightly
viscous rotating fluids contained in spherical shells. For very small
viscosities, relevant to astrophysical and geophysical applications,
oscillating flows are confined within shear layers that closely follow
attractors of characteristics. In the two-dimensional limit of a
spherical shell, \cite{RVG02} derived analytical solutions for the
structure of shear layers. They showed in particular that the velocity
field of eigenmodes obeys the same equation as that governing the
quantum states of a particle trapped in a parabolic well (the
Schrödinger equation).

More recently, forced oscillations in rotating fluids have been
examined in the context of tidal interactions between two celestial
bodies \citep{OL04, WU05b, OL07, GL09, O09, RV10}.  Dissipation of
tidally-forced modes of oscillations in each companion alters the
orbital elements and spin of each body. It plays a prominent role in
the late evolution of close-in extrasolar planets. The efficiency of
tidal dissipation mechanisms is usually parametrized by a single,
dimensionless tidal quality factor $Q$, which is an inverse measure of
the tidal torque acting on a gravitationally perturbed body. However,
tidal dissipation is extremely sensitive to the frequency and
amplitude of the external tidal potential \citep[see, e.g.,][]{O09,
  RV10}. This dependency is intimately related to the existence of
singularities arising from the hyperbolic nature of the Poincar{\'e}
equation. As in free modes of oscillations, these singularities
consist of the critical latitude singularity and attractors of
characteristics.

Works examining tidal dissipation mechanisms in rotating fluids have
used simple fluid models, where the background flow disturbed by
inertial modes is in pure solid-body rotation.  The rotation pattern
inside stars and planets is much more intricate, as may be
guessed. Large-scale flows are usually present in stars and
planets. Most frequently, such flows are largely axisymmetric, and
include differential rotation as well as a weak meridional
circulation. In light of previous works on rigidly rotating fluids,
the present paper is a first study of how departure to solid-body
rotation impacts the oscillation properties of rotating fluids and
their associated tidal dissipation. In this first paper, we focus on
free inertial modes of oscillations. Tidally-forced inertial modes
will be examined in a future study. As a first simplification, we
discard meridional circulation, which is very weak in general,
especially for very small viscosities. However, the choice for the
differential rotation pattern, which we assume to be axisymmetric,
remains open. The {\color{black} azimuthal velocity of the} (smooth)
differentially rotating background flow can be expanded into spherical
harmonics,

\[ u_\varphi(r, \theta) = - \sum_\ell \wlm(r)\partial_\theta\YL\; ,\]
as for any axisymmetric azimuthal component of a vector field
\cite[][]{Rieutord87}. In this expression, $r$, $\theta$, and
$\varphi$ denote the usual spherical coordinates, {\color{black}
  $Y_{\ell}^m$ the spherical harmonics normalized on the sphere of
  unit radius, and $\wlm(r)$ accounts for the radial variation in the
  flow.}  If differential rotation is smooth, we expect the previous
expansion to be dominated by small $\ell$'s. In the present work, we
will primarily focus on a single term, the $\ell$=1-term, such that
the background velocity field takes the form
\[ u_{\varphi}(r, \theta) = w(r)\sth = r\Omega(r)\sth,\] with
differential rotation along the radial direction only {\color{black}
  ($\Omega$ denotes the local angular frequency of the fluid)}. The
obvious asset of this functional form is that it preserves the
relative simplicity of the coupling of harmonics imposed by the
Coriolis term. This type of flow is known in the astrophysical
literature as \emph{shellular} differential rotation
\cite[][]{zahn92}. It originates from a simple description of the
baroclinic flow (the thermal wind in geophysics), which arises in
stably stratified radiative zones of rotating stars \cite[see,
e.g.,][]{R06}. The rotation is assumed to be uniform on spheres as a
consequence of some shear-induced turbulence, which erases latitudinal
gradients of the velocity \cite[][]{zahn92}. Low-frequency
perturbations of such flows are driven by the combined effects of the
Coriolis acceleration and buoyancy. Assuming differential rotation
along the radial direction therefore comes to neglecting the effects
of buoyancy on the fluid perturbations.  Our model approaches this
astrophysical configuration in the limit of weak stratification, when
the \BVF\ is small compared to the Coriolis frequency. When this is
the case, some modes, called H1 in \cite{Dintrans99}, are essentially
inertial modes slightly perturbed by buoyancy (i.e., the Coriolis
acceleration is the main restoring force).  In addition to giving a
first look at the effect of differential rotation on inertial
oscillations, shellular rotation may also provide a (partial) view on
the properties of waves propagating in radiative zones inside rotating
stars.

To complement the study of shellular flows, which may mimic stellar
baroclinic flows, we also consider differential rotation associated
with a barotropic fluid model. In this case, differential rotation
only depends on the cylindrical radial coordinate $s =
r\sth$. Cylindrical differential rotation is expected in the
convective zones of rapidly rotating stars or giant planets. It has
been much investigated in the context of the Taylor-Couette
instability, but in cylindrical geometry. However, recent experimental
work by \cite{Kelley2007, KTZL10} and \cite{RTZL12} has raised the
question of how the properties of inertial waves in spherical geometry
are changed by the inclusion of cylindrical rotation.  Since spherical
geometry combined with cylindrical rotation results in the variables
non-separability, which is a necessary condition for the existence of
singularities like attractors, some properties of the perturbations
can be derived by the sole analysis of the dynamics of
characteristics.

This paper is organized as follows. Our physical model is described in
\S~\ref{sec:lin}. The ideal case of an inviscid differentially
rotating fluid is first examined in \S~\ref{sec:inv} through a linear
analysis in the short-wavelength approximation. Cylindrical and
shellular rotation profiles are considered in \S~\ref{sec:inv_cyl}
and~\ref{sec:inv_shell}, respectively. A detailed study of the
dynamics of characteristics allows us to determine the range of
frequencies at which inertial oscillations may {\color{black}exist} in
differentially rotating fluids. The presence of turning surfaces,
critical latitudes, and corotation resonances within the fluid is
investigated. The full problem of a viscously rotating fluid is then
investigated in \S~\ref{sec:shear} with (linear) numerical
calculations. Again, results with cylindrical and shellular rotation
profiles are presented. Shear layers are shown to closely follow the
attractors of characteristics obtained for an inviscid fluid in the
short-wavelength approximation. Our simulations primarily focus on
axisymmetric modes (\S~\ref{sec:num_axi}), but a few examples of
non-axisymmetric modes with a corotation resonance are also examined
(\S~\ref{sec:num_nonaxi}).  Concluding remarks and future directions
are drawn in \S~\ref{sec:discu}.

\section{Inertial modes in a differentially rotating shell: physical model}
\label{sec:lin}

We consider a differentially rotating viscous fluid inside a spherical
shell, located between radii $\eta R$ and $R$ ($0 < \eta < 1$). The
fluid is assumed to be {\color{black}homogeneous}, incompressible, and
of constant kinematic viscosity $\nu$. We denote by $\vect{u}$ the
fluid's velocity.  Any quantity $x$ is decomposed as $x=x_0 + x_1$,
where $x_0$ describes an unperturbed background quantity and $x_1$ the
associated disturbance ($|x_1| \ll |x_0|$). In an inertial frame, and
using the set of spherical coordinates $(r,\theta,\varphi)$, the
linearized Navier-Stokes equation reads

\begin{equation}
\frac{\partial \vect{u_1}}{\partial t} 
+ 
\Omega_0 \frac{\partial \vect{u_1}}{\partial \varphi} 
+
2\Omega_0 \vect{e_z} \times \vect{u_1}
+
A(\vect{u_1}) \vect{e_{\varphi}}
= 
-\vect{\nabla} p_1 + \nu \vect{\Delta u_1},
\label{eq:mom}
\end{equation}
with $\Omega_0 = u^{\varphi}_0 / r\sin\theta$ the fluid's angular
velocity, $u^{\varphi}_0 = \vect{u_0}\cdot\vect{e_\varphi}$,
$\vect{e_z}=\cos\theta\vect{e_r} - \sin\theta\vect{e_\theta}$ is the
unit vector along the rotation axis, {\color{black}and $p_1$ denotes
  the pressure perturbation divided by the reference density}. Since
the fluid is incompressible, the linearized continuity equation reads
\begin{equation}
\vect{\nabla}\cdot\vect{u_1} = 0.
\label{eq:div}
\end{equation}
In Eq.~(\ref{eq:mom}), $A(\vect{u_1})$ is proportional to the shear in 
the background flow:
\begin{equation}
A(\vect{u_1}) = r \sin\theta \;\vect{u_1} \cdot \vect{\nabla}\Omega_0.
\label{eq:Av1}
\end{equation}
In the following we consider either
\begin{itemize}
\item A cylindrical rotation profile, $\Omega_0 = \Omega_0(s)$, with
  $s = r\sin\theta$ the radial cylindrical coordinate. This rotation
  profile satisfies the Proudman-Taylor theorem for an incompressible
  {\color{black}homogeneous fluid in the inviscid limit}, and no 
  additional body-force needs to be assumed to ensure an
  equilibrium state for the differentially rotating background flow.
\item A shellular rotation profile, $\Omega_0 = \Omega_0(r)$. In this
  case, differential rotation is assumed to be maintained by some
  appropriate body-force $\vect{f}$ (arising for instance from
  Reynolds stresses {\color{black}or from a latitudinal entropy
    gradient}), not included in our model. {\color{black}This
    body-force should verify $\vect{\nabla} \times \vect{f} =
    -s \partial\Omega^2_0 / \partial z\,\vect{e_{\varphi}}$ for
    Taylor-Proudman theorem to be satisfied in this case.}
\end{itemize}

We seek solutions {\color{black} to} Eqs.~(\ref{eq:mom})
and~(\ref{eq:div}) in the form $\exp(i\Omega_p t + im\varphi)$, where
$\Omega_p$ denotes the wave angular frequency in the inertial
frame. {\color{black}We introduce the Ekman number $E = \nu / R^2
  \Omega_{\rm ref}$, where the reference angular frequency,
  $\Omega_{\rm ref}$, refers to $\Omega_0(R)$ for shellular rotation
  or to $\Omega_0(s=0)$ for cylindrical rotation}.  Dropping all
subscripts and taking the curl of Eq.~(\ref{eq:mom}), we are to solve:
\greq \vect{\nabla} \times \left( i\tilde{\Omega}_p \vect{u} + 2\Omega
  \vect{e_z} \times \vect{u} + r \sin\theta (\vect{u} \cdot
  \vect{\nabla}\Omega) \vect{e_{\varphi}} \right)
= E \vect{\nabla} \times\Delta\vect{u}\\
\\
\vect{\nabla}\cdot\vect{u}=0, 
\egreqn{eq:vort}
where $\tilde\Omega_p = \Omega_p + m\Omega$ is the Doppler-shifted
wave frequency (that is, the wave frequency {\color{black} in} the
frame rotating with the fluid).

We complete Eqs.~\eq{eq:vort} with stress-free boundary conditions at
both boundaries: $\vect{u} \cdot \vect{e_r} = 0$ and $\vect{e_r}\times
[\sigma] \vect{e_r} = {\bf 0}$, where $[\sigma]$ denotes the viscous
stress tensor. As discussed in \cite{RV10}, the general properties of
oscillation modes are essentially unchanged by the choice of boundary
conditions. We thus opt for stress-free boundaries, which are
numerically less demanding.

\section{Inviscid problem: paths of characteristics, existence of
  turning surfaces and corotation resonances}
\label{sec:inv}
Before examining the full solutions to Eqs.~\eq{eq:vort} by means of
numerical simulations, we first study the inviscid limit and
especially the dynamics of characteristics, which is known to feature
the eigenfunctions of the viscous problem. For this purpose, we shall
use cylindrical coordinates $(s,\varphi,z)$, and eliminate the
velocity perturbations in favor of the pressure perturbations
($p$). {\color{black} The partial differential equation (PDE) satisfied
  by $p$ is given in the appendix by
  Eq.~(\ref{eq_poin_app}). Retaining only the second-order terms,
  Eq.~(\ref{eq_poin_app}) becomes}:
\begin{equation}
  \frac{\partial^2 p}{\partial s^2} +
  \frac{A_z}{\tilde\Omega^2_p} \frac{\partial^2 p}{\partial
    s \partial z} + \left( 1 - \frac{A_s}{\tilde\Omega^2_p}
  \right) \frac{\partial^2 p}{\partial z^2} = 0,
\label{poincare2}
\end{equation} 
where 
\begin{equation} A_s = \frac{2\Omega}{s}\frac{\partial}{\partial s}
  \left( s^2 \Omega \right) \;\;\;\;{\rm and} \;\;\;\; A_z =
  \frac{2\Omega}{s}\frac{\partial}{\partial z} \left( s^2 \Omega
  \right).
\label{eq_cylkappas} 
\end{equation}
Eq.~(\ref{poincare2}) matches equation (A 16) of \cite{Ogilvie05} for
{\color{black}an homogeneous} incompressible {\color{black}fluid}.
{\color{black} The equation governing the paths of characteristics in a
  meridional plane then reads
\begin{equation}
\frac{dz}{ds} = \frac{1}{2\tilde\Omega^2_p} \left(
A_z \pm \xi^{1/2}
\right)
\;\;\;\;{\rm with}\;\;\;\;
\xi(s,z) = A_z^2 + 4\tilde\Omega^2_p (A_s - \tilde\Omega^2_p).
\label{eq_cargeneral} 
\end{equation}
For solid-body rotation, $A_z = 0$, $A_s = 4\Omega^2$, and
Eq.~(\ref{poincare2}) reduces to the well-known Poincar{\'e} equation
for inertial waves. In this case, characteristics form straight lines
in a meridional plane, and the fact that $\xi$ is constant makes the
entire shell either hyperbolic or elliptic for inertial waves.  With
differential rotation, however, the dependence of $A_s$, $A_z$ and
$\Omega$ with $s$ and $z$ implies that characteristics no longer form
straight lines, as will be illustrated in the following. Also, $\xi$
may vanish and change sign inside the shell. This makes possible the
co-existence of hyperbolic and elliptic domains for pure inertial
waves in a differentially rotating spherical shell, separated by
turning surfaces. It is reminiscent of the existence of turning
surfaces for gravito-inertial waves and magneto-inertial waves in
rigidly rotating spherical shells, for which the perturbed pressure
also satisfies a second-order PDE of mixed type in the inviscid limit
\citep{FS82b, Fried82, Fried87, Dintrans99}.
Equation~(\ref{eq_cargeneral}) also shows that, in contrast to
solid-body rotation, paths of characteristics with differential
rotation depend on the azimuthal wavenumber $m$ through
$\tilde\Omega_p$.}  In {\color{black} the rest of} this section, we
study {\color{black} in more detail} the dynamics of characteristics
based on Eq.~(\ref{eq_cargeneral}), considering {\color{black} a
  particular profile of} cylindrical rotation in \S~\ref{sec:inv_cyl},
and of shellular rotation in \S~\ref{sec:inv_shell}.

\subsection{Cylindrical rotation}
\label{sec:inv_cyl}

\subsubsection{Paths of characteristics}
To restrict the differential rotation function space to a
one-parameter space, we consider the following background rotation
profile: \beq \Omega(s) / \Omega_{\rm ref} = 1 + \varepsilon (s/R)^2,
\eeq {\color{black}with $\Omega_{\rm ref} = \Omega(s=0)$ the shell's
  angular frequency at the rotation axis, and where $|\varepsilon|$
  can be seen as a Rossby number for the fluid.}
Eq.~(\ref{poincare2}) then takes the form:
\begin{equation}
  \frac{\partial^2 p}{\partial s^2} 
  + \left( 1 - \frac{A_s}{\tilde\Omega^2_p} \right) 
  \frac{\partial^2 p}{\partial z^2} = 0,
	\label{cyl_poincare}
\end{equation}
{\color{black}where}
\begin{equation}
  \tilde\Omega_p(s) = \Omega_p + m\Omega(s)
\label{cyl_tildeomega}
\end{equation}
{\color{black}is the Doppler-shifted frequency,} and
\begin{equation}
  A_s(s)
  =
  4\Omega^2(s) \times \left[ 1 + \frac{\varepsilon(s/R)^2}{1 + \varepsilon (s/R)^2} \right].
\label{cyl_kappas}
\end{equation}
We will take $\varepsilon > -1/2$ so that {\color{black} $A_s >
  0\;\forall s \in [0,R]$ (Rayleigh's stability criterion). We will
  denote the quantity $A_s$ by $\kappa^2_s$ (it is known as the square
  of the radial epicyclic frequency in astrophysical discs).  We point
  out that Eq.~(\ref{cyl_poincare}) is symmetric by $z \rightarrow
  -z$, so that only positive values of $z$ will be considered. From
  Eq.~(\ref{eq_cargeneral}), paths of characteristics satisfy}
\begin{equation}
  \frac{dz}{ds} = \pm \xi^{1/2},
  \label{cyl_caract}
\end{equation}
with
\begin{equation}
  \xi(s) = \frac{\kappa^2_s(s)}{\tilde\Omega^2_p(s)}-1.
  \label{cyl_xi}
\end{equation}

\subsubsection{Existence of turning surfaces: D and DT inertial modes}
\label{sec:cyl_DDT}
The relation $\xi(s) = 0$ defines turning surfaces that {\color{black}
  separate the hyperbolic and elliptic domains. They} correspond to
cylinders in a meridional slice of the shell. Their location satisfies
\begin{equation}
  \tilde\Omega^2_p(s) = \kappa^2_s(s),
  \label{eq:cyl_turning}
\end{equation}
with waves propagating when $\kappa^2_s(s) \geq
\tilde\Omega^2_p(s)$. {\color{black} Here and in the following, we will
  assume that the hyperbolic domains of the spherical shell, where
  $\xi$ is positive, host inertial modes of oscillations.} The
dependence of $\xi$ {\color{black} on} $s$ leads to two kinds of
inertial modes with cylindrical rotation:
\begin{enumerate}
\item[(i)] modes that exhibit at least one turning surface within the
  shell, which we call DT modes for future reference (D for
  differential rotation, and T for turning surface),
\item[(ii)] generalized inertial modes with no turning surfaces within
  the shell, which we name D modes.
\end{enumerate}

\begin{figure}
  \centering
  \resizebox{\hsize}{!}
  {
    \includegraphics{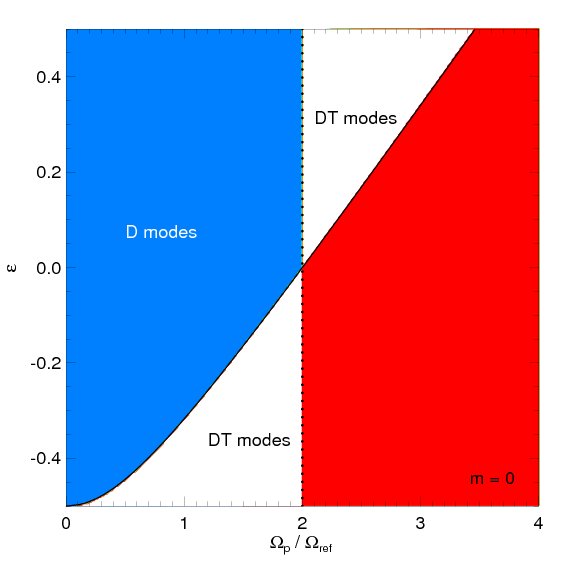}
    \includegraphics{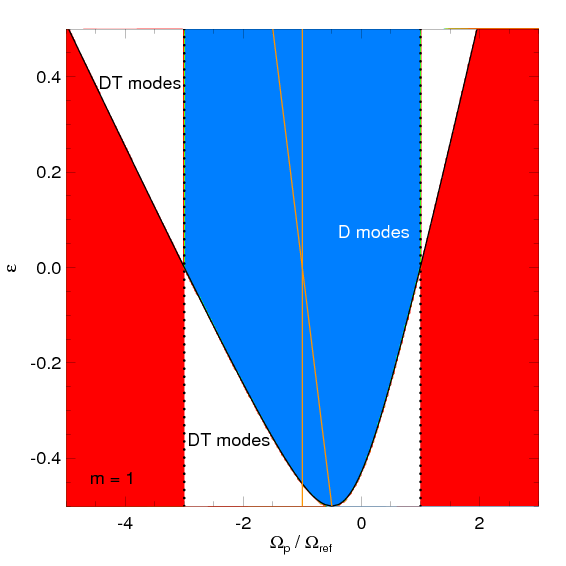}
  }
   \resizebox{\hsize}{!}
  {
    \includegraphics{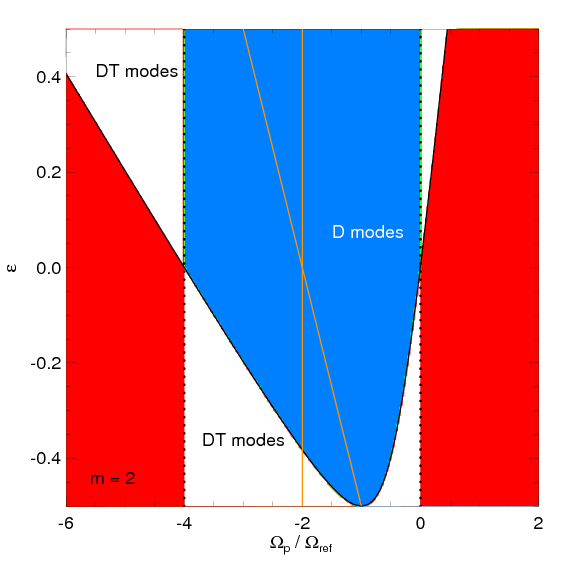}
    \includegraphics{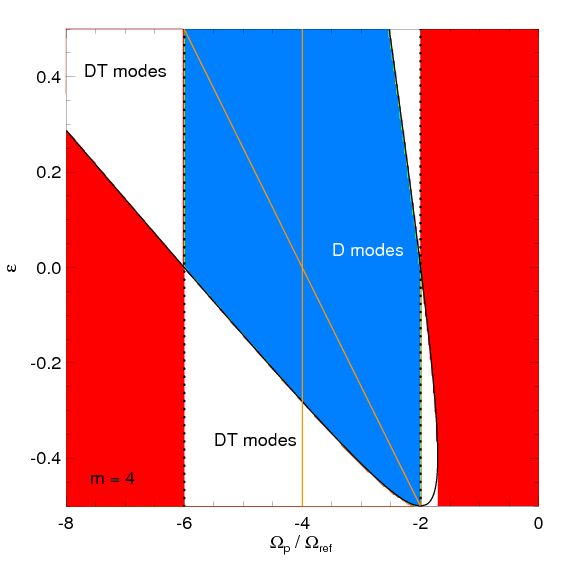}
  }
  \caption{\label{cyl_turning}Illustration of the two kinds of
    inertial modes propagating in a differentially rotating fluid,
    with cylindrical rotation profile $\Omega(s) / \Omega_{\rm ref} =
    1 + \varepsilon (s/R)^2$ {\color{black}($\Omega_{\rm ref}$ denotes
      the shell's angular frequency at the rotation axis)}.  Results
    are shown for azimuthal wavenumbers $m$ equal to 0, 1, 2 and 4
    from top-left to bottom-right. DT modes, represented by white
    areas, exhibit turning surfaces inside the shell. D modes
    correspond to the blue area and do not feature turning surfaces
    inside the shell. No inertial modes may propagate in the red
    areas.  The solutions to $\xi(s=0)=0$ are shown by vertical dotted
    lines, and the solutions to $\xi(s=R)=0$ by solid black curves.
    Non-axisymmetric modes that feature a critical cylinder inside the
    shell {\color{black}(corotation resonance)} are
    {\color{black}located between} the two {\color{black} orange solid}
    curves.}
\end{figure}
The occurrence of axisymmetric ($m=0$) and of some non-axisymmetric D
and DT modes is illustrated in Fig.~\ref{cyl_turning}. In all panels,
the mode's eigenfrequency in the inertial frame ($\Omega_p$) is shown
in x-axis, and the differential rotation parameter $\varepsilon$ is in
y-axis. Contours are obtained by numerically calculating the
right-hand side of Eq.~(\ref{cyl_xi}).  DT modes, which are
represented by white areas, have $\xi < 0$ at least once within the
shell. D modes satisfy $\xi \geq 0$ everywhere in the shell and are
shown by the blue area. No inertial modes can propagate if $\xi < 0$
in the entire shell, such case being depicted in
red. {\color{black}Note} that axisymmetric inertial waves may propagate
for eigenfrequencies $\Omega_p > 2\Omega_{\rm ref}$ if $\varepsilon >
0$ (see top-left panel in Fig.~\ref{cyl_turning}). {\color{black} Note
  also that the results shown in Fig.~\ref{cyl_turning} are
  independent of the shell's aspect ratio $\eta$, as is actually the
  dynamics of characteristics with our cylindrical rotation profile.}

The transition curves that separate D and DT modes satisfy
$\xi(s=0)=0$ and $\xi(s=R)=0$. The solution to $\xi(s)=0$ can be cast
as $\Omega_p / \Omega_{\rm ref} = -m[1+\varepsilon (s/R)^2] \pm
2\sqrt{[1+2\varepsilon (s/R)^2][1+\varepsilon (s/R)^2]}$. Applied to
the following cases:
\begin{itemize}
\item $\xi(s=0)=0$, that is when the turning surface matches the
  location of the {\color{black}rotation} axis, we have
\begin{equation}
	\Omega_p / \Omega_{\rm ref}  = -m \pm 2,
	\label{cyl_xis0}
\end{equation}
such case being depicted by vertical dotted lines in
Fig.~\ref{cyl_turning}, \smallskip
\item $\xi(s=R)=0$, when the turning surface passes by the shell's 
surface, yields
\begin{equation}
	\Omega_p / \Omega_{\rm ref} = -m(1+\varepsilon) \pm 2\sqrt{(1+2\varepsilon)(1+\varepsilon)},
	\label{cyl_xis1}
\end{equation}
which is overplotted by solid curves in Fig.~\ref{cyl_turning}.
\end{itemize}
{\color{black} We point out that there may exist two turning surfaces
  in a meridional plane. This occurs for instance for the DT modes
  with $m=4$ located in the (small) region of parameter space below
  the solid curve near $\Omega_p = -2 \Omega_{\rm ref}$ and
  $\varepsilon = -0.5$.}

Non-axisymmetric modes may feature a corotation resonance within the
shell when $\tilde\Omega_p(s) = 0$. The location of such
\emph{critical cylinders} is given by
\begin{equation}
  s_{\rm crit} = R\,\sqrt{ \varepsilon^{-1} \times \left( -1 - \frac{\Omega_p}{m\Omega_{\rm ref}}\right) }.
  \label{scrit}
\end{equation}
For $m > 0$, modes with $\varepsilon \leq 0$ have a critical layer if
$-m \leq \Omega_p / \Omega_{\rm ref} \leq -m(1+\varepsilon)$, and
those with $\varepsilon \geq 0$ if $-m(1+\varepsilon) \leq \Omega_p /
\Omega_{\rm ref} \leq -m$.  Modes that exhibit a critical cylinder
within the shell are {\color{black} located between the two orange
  solid} curves in Fig.~\ref{cyl_turning}.

\begin{figure}
  \centering
  \resizebox{0.5\hsize}{!}
  {
   \includegraphics{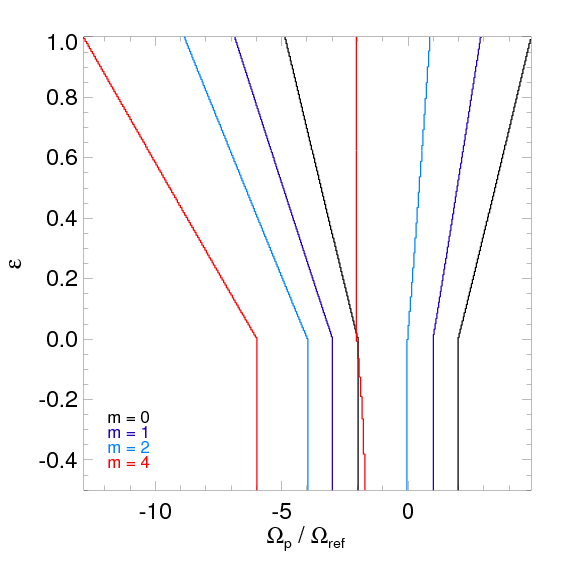}
  }   
  \caption{\label{cyl_turning2}Range of eigenfrequencies ($\Omega_p$,
    x-axis) for inertial modes of oscillations in a rotating shell
    with cylindrical rotation profile $\Omega(s) / \Omega_{\rm ref} =
    1 + \varepsilon (s/R)^2$. {\color{black} Eigenfrequencies are in
      units of $\Omega_{\rm ref}$, the shell's angular frequency at
      the rotation axis}. Results are for $m=0$, 1, 2 and 4.  For a
    given $m$, the minimum and maximum frequencies at which inertial
    modes {\color{black}exist} correspond to the left and right curves,
    respectively.}
\end{figure}
The minimum and maximum frequencies at which inertial waves may
propagate in a shell with cylindrical rotation is summarized in
Fig.~\ref{cyl_turning2} for the four azimuthal wavenumbers considered
in Fig.~\ref{cyl_turning}. These frequencies are obtained by solving
$\xi(s) = 0$. They correspond to the minimum and maximum frequencies
amongst the four values given by Eqs.~(\ref{cyl_xis0})
and~(\ref{cyl_xis1}), except for $m=4$ and $\varepsilon$ approaching
$-0.5$ (in which case, as already pointed out before, there are two
turning surfaces inside the shell).  Fig.~\ref{cyl_turning2}
highlights that the range of eigenfrequencies at which inertial modes
of oscillations may {\color{black}exist} in a spherical shell is
broader with cylindrical rotation than with solid rotation,
particularly \emph{if} the rotation profile is an increasing function
of the cylindrical radial coordinate. This is expected, because of the
larger frequencies at which the fluid may rotate in this case.

\subsubsection{Dispersion relation, phase and group velocities}
We examine in this paragraph the propagation properties of local waves
in a meridional plane. {\color{black}In the short-wavelength
  approximation, the general expressions for the wave dispersion
  relation, the phase and group velocities are given in the Appendix
  by Eqs.~(\ref{disp_app}), ~(\ref{vphase_app}) and
  ~(\ref{vgroup_app}).  For our cylindrical rotation profile,
  Eq.~(\ref{disp_app}) reduces to}
\begin{equation}
	\tilde\Omega_p^2 (s) = \kappa_s^2(s) \frac{k_z^2}{ \|{\bf k}\|^2 }, 
\label{cyl_disp1}
\end{equation}
where $\|{\bf k}\| = \sqrt{k_s^2 + k_z^2}$. In the frame rotating 
with the fluid, the phase velocity, ${\bf v_p} = \tilde\Omega_p {\bf
  k} / \|{\bf k}\|^2$, satisfies
\begin{equation}
	{\bf v_p} = \pm \kappa_s(s) \frac{k_z {\bf k}}{\|{\bf k}\|^3},
	\label{cyl_vphase}
\end{equation}
while the group velocity, ${\bf v_g} = \partial \tilde\Omega_p
/ \partial {\bf k}$, is given by
\begin{equation}
  {\bf v_g} = \pm \kappa_s(s) \frac{k_s}{\|{\bf k}\|^3} (-k_z {\bf e_s} + k_s {\bf e_z}).
	\label{cyl_vgroup}
\end{equation}
From Eqs.~(\ref{cyl_xi}) and~(\ref{cyl_disp1}), $k_s = 0$ at a turning
surface (where $\xi$ vanishes). At this location, the radial component
of the phase velocity therefore vanishes, as well as the two
{\color{black} other} components of the group velocity (that is, rays
are perpendicular to the turning surface).

A corotation resonance occurs inside the shell where $\tilde\Omega_p
(s) = 0$. This formally happens if $|k_s| \rightarrow \infty$, in
which case the modulus of the phase velocity and the group velocity
both tend to zero, or if $k_z \rightarrow 0$, in which case the phase
velocity cancels out while the radial component of the group velocity
cancels out. In both cases, characteristics become more and more
parallel to the rotation axis as they approach the corotation
resonance (their slope $|dz/ds|$ increases), and they do not cross
corotation (as the radial component of the group velocity tends to
zero at this location). The fact that the phase velocity progressively
decreases to zero upon approaching corotation suggests that non-linear
effects should become important near this location, as pointed out for
instance by \cite{BO10} for internal gravity waves {\color{black} in a
  background shear flow, which satisfy a dispersion relation similar
  to that in Eq.~(\ref{cyl_disp1}) with the epicyclic frequency
  squared $\kappa_s^2$ replaced by the \BVF squared.}

\subsection{Shellular rotation}
\label{sec:inv_shell}
We now come to the propagation properties of inertial modes of
oscillation in fluids with shellular rotation (the background rotation
profile being a prescribed function of the spherical radius). As in
the case of cylindrical rotation, we show in this section that two
kinds of inertial modes can be distinguished, based on the presence or
not of turning surfaces within the shell. We then discuss the range of
eigenfrequencies at which inertial modes may exist, the existence of
critical latitudes and corotation resonances.

\subsubsection{D and DT inertial modes}
\label{sec:poincare}
We adopt the same strategy as in \S~\ref{sec:inv_cyl} for cylindrical
rotation and restrict the differential rotation function space to a
one-parameter space. We take the following shellular rotation profile:
\beq \Omega(r) = \Omega_{\rm ref} \times (r/R)^\sigma, \eeq
{\color{black} where $r = \sqrt{s^2 + z^2}$}, $\Omega_{\rm ref} =
\Omega(r=R)$ is the angular frequency at the shell's surface, and
$\sigma > -2$ to avoid centrifugal instabilities
{\color{black}(Rayleigh's stability criterion, $A_s > 0$, is
  satisfied). The quantity $|1-\eta^{\sigma}|$, which measures the
  relative difference of angular frequencies between the inner core
  and the shell's surface, can be seen as a Rossby number for the
  fluid}. Eq.~(\ref{poincare2}) being symmetric by $z \rightarrow -z$
for shellular rotation profile, only positive values of $z$ will be
considered (as with cylindrical rotation). From
Eq.~(\ref{eq_cargeneral}), paths of characteristics satisfy
\begin{equation}
	\frac{dz}{ds} = A(r) \frac{sz}{R^2} \pm \xi^{1/2}, \label{caract}
\end{equation} 
where
\begin{equation}
  \xi(s,z) = A^2(r) \frac{s^2z^2}{R^2} + 2A(r)\frac{s^2}{R^2} + \frac{4\Omega^2(r) -
    \tilde\Omega^2_p(r)}{\tilde\Omega^2_p(r)}, \label{xi}
\end{equation}
\begin{equation}
	A(r) = \sigma \, \frac{R^2}{r^2} \,
	\frac{\Omega^2(r)}{\tilde\Omega^2_p(r)},  \label{Ax}
\end{equation}
and $\tilde\Omega_p(r) = \Omega_p + m\Omega(r)$ is the Doppler-shifted
wave frequency. Whenever paths of characteristics are calculated by
solving Eq.~(\ref{caract}) (or Eq.~(\ref{cyl_caract}) for cylindrical
rotation), reflections are imposed at the shell's inner core ($r=\eta
R$) and surface ($r=R$), at the rotation ($s=0$) and equatorial
($z=0$) axes, and at the location of turning surfaces (where $\xi=0$).

\begin{figure}
  \centering
  \resizebox{\hsize}{!}
  {
    \includegraphics{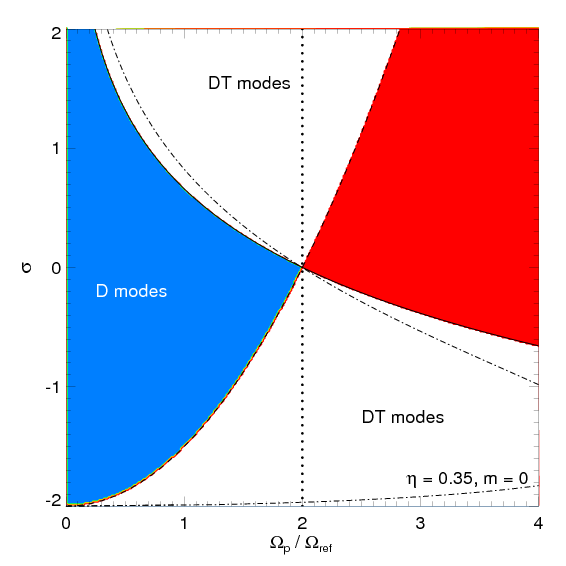}
    \includegraphics{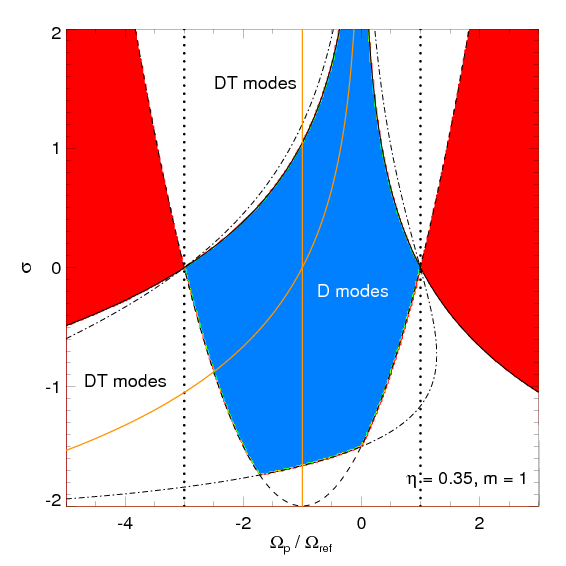}
  }
   \resizebox{\hsize}{!}
  {
    \includegraphics{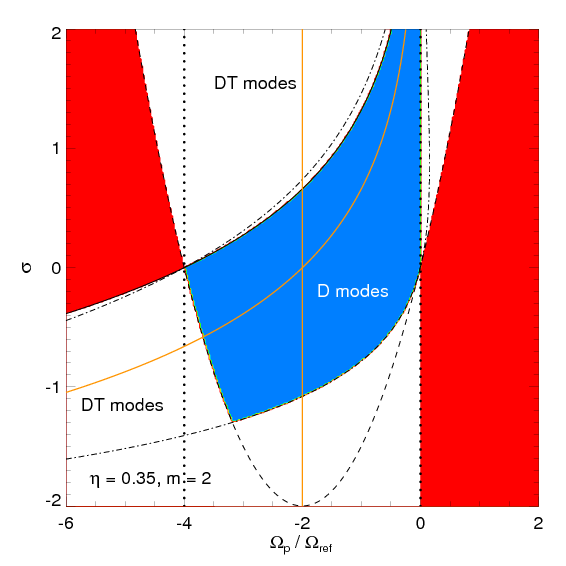}
    \includegraphics{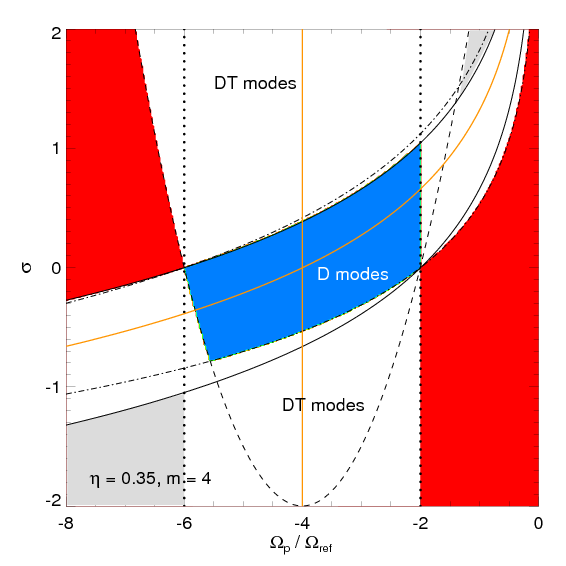}
  }
  \caption{\label{turning}Occurrence of D and DT inertial modes with
    {\color{black}the shellular rotation profile $\Omega(r) /
      \Omega_{\rm ref} = (r/R)^{\sigma}$ ($\Omega_{\rm ref}$ denotes
      the angular frequency at the shell's surface).} Results are
    obtained for $\eta=0.35$, and $m=$ 0, 1, 2 and 4 from top-left to
    bottom-right. DT modes (white areas) have at least one turning
    surface inside the shell. D modes (blue area) do not have turning
    surfaces. No inertial modes exist within the red areas. The
    dotted, dashed, solid and dash-dotted curves are solutions to
    Eqs.~(\ref{eq:dwmodes_dotted}) to~(\ref{eq:border4}),
    respectively. In the bottom-right panel, grey areas depict DT
    modes located between two turning surfaces that englobe the
    shell's inner core and surface. {\color{black} For $m \neq 0$,
      modes that feature a critical layer inside the shell (where
      $\tilde\Omega_p(r) = 0$) are located between the two orange
      solid curves.}  }
\end{figure}

The possibility that $\xi$ vanishes within the shell leads again to
two kinds of inertial modes with shellular rotation: DT modes, with at
least one turning surface inside the shell, and D modes, with no
turning surfaces inside the shell.  The occurrence of axisymmetric and
of some non-axisymmetric D and DT modes is depicted in
Fig.~\ref{turning}. Contours are obtained upon calculation of
Eq.~(\ref{xi}) in a spherical shell with $\eta=0.35$. As in
Fig.~\ref{cyl_turning}, D and DT modes are depicted by blue and white
areas, respectively, and no inertial modes exist in the red areas.

The relation $\xi = 0$ defines turning surfaces with colatitude
$\theta(r)$ given by
\begin{equation}
  \sin^2\theta (r) = \frac{1}{2} \left[ 1 + \frac{2\tilde\Omega_p^2(r)}{\sigma \Omega^2(r)} \pm \mu^{1/2}(r) \right],
	\label{eq:turning}
\end{equation}
with
\begin{equation}
	\mu(r) = 1 + \frac{4(4+\sigma)\tilde\Omega_p^2(r)}{\sigma^2 \Omega^2(r)}.
\end{equation}
Another way to find the occurrence of DT modes is to require that, for
given values of $m$, $\Omega_p$ and $\sigma$, the inequality $0 \leq
\sin^2\theta(r) \leq 1$ is satisfied at least once within the
shell. We also point out that the two propositions $\forall \{s,z\}$,
$\xi(s,z) > 0$, and $\forall r$, $|\tilde\Omega_p(r)| < 2\Omega(r)$,
seem to be equivalent numerically (as we have checked).  In other
words, if the whole shell satisfies the condition for hyperbolicity,
it also meets the standard criterion for propagation of inertial waves
everywhere locally. Thus, regions with no mode propagation, and those
with D modes, can be distinguished by requiring that D modes verify
$|\tilde\Omega_p(r)| < 2\Omega(r)$ in the whole shell.

The transition curves that separate D and DT modes then satisfy:
\begin{itemize}
\item $\sin^2 \theta(R)=0$, when the turning surface hits the
  shell's surface at colatitude $\theta = 0$. It reads
  $|\tilde\Omega_p(R)| = 2\Omega(R)$, or
  \begin{equation}
   \Omega_p / \Omega_{\rm ref} = -m \pm 2, 
  \label{eq:dwmodes_dotted}
  \end{equation}
  expression that is depicted by vertical dotted lines in
  Fig.~\ref{turning}.  \smallskip
\item $\sin^2 \theta(R)=1$, which yields 
\begin{equation}
	{\color{black}
	\Omega_p  / \Omega_{\rm ref} = -m \pm \sqrt{2(2+\sigma)},
	}
	\label{eq:dwmodes}
\end{equation}
which is overplotted by a dashed curve.
\smallskip
\item $\sin^2 \theta(\eta R)=0$, which can be recast as
\greq 
 {\color{black}
 \Omega_p  / \Omega_{\rm ref} = (2-m)\eta^{\sigma}
 }
 \qquad
\textrm{\hspace{4mm}if\; $\Omega_p / (2-m) > 0$}\\
\\
{\color{black} \Omega_p / \Omega_{\rm ref} = -(2+m)\eta^{\sigma} }
\qquad \textrm{\hspace{1mm}if\; $\Omega_p / (2+m) < 0$},
\egreqn{eq:dwmodes2} and are shown by solid lines.
\smallskip
{\color{black} 
\item $\sin^2 \theta(\eta R)=1$, when the turning surface hits the
  shell's inner core at colatitude $\theta=\pi/2$, that is when
\begin{equation}
  \Omega_p  / \Omega_{\rm ref} = \eta^{\sigma} \left( -m \pm 2\sqrt{1+\sigma/2}\right),
	\label{eq:border4}
\end{equation}
expression that is shown by dash-dotted curves.
} 
\end{itemize}

\begin{figure}
  \centering
  \resizebox{0.5\hsize}{!}
  {
   \includegraphics{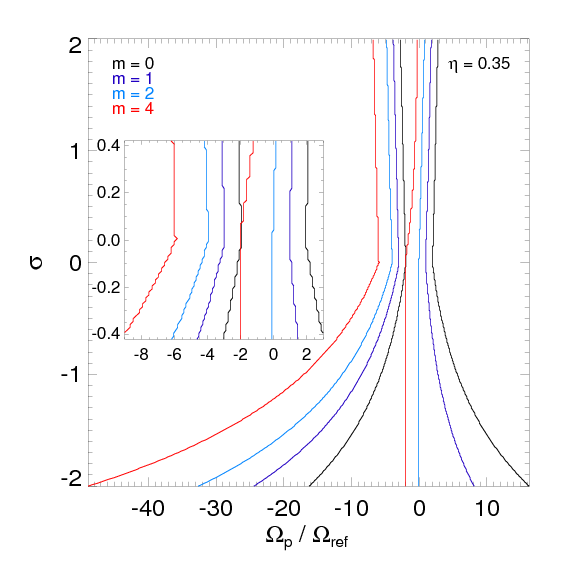}
  }   
  \caption{\label{turning2}Minimum and maximum eigenfrequencies
    (x-axis) at which inertial modes of oscillation may
    {\color{black}exist} with shellular rotation profile $\Omega(r) /
    \Omega_{\rm ref} = (r/R)^{\sigma}$, for $m=0$, 1, 2 and 4
    ($\eta=0.35$). {\color{black}Eigenfrequencies are in units of the
      angular frequency at the shell's surface}. The inset plot
    focuses on previous range of eigenfrequencies for $\sigma \in
    [-0.4,0.4]$.}
\end{figure}
DT modes with $|m| \leq 2$ are located between the turning surface and
the shell's surface for $\sigma > 0$ (since the background velocity
increases with radius), and between the shell's inner core and the
turning surface for $\sigma < 0$. A more complex situation arises when
$|m| > 2$, where the equation $|\tilde\Omega_p(\eta R)| = 2\Omega(\eta
R)$ may admit two solutions for a given $\Omega_p$, as can be seen
from Eq.~(\ref{eq:dwmodes2}). In the bottom-right panel of
Fig.~\ref{turning} ($m=4$), DT modes located between the two solid
lines propagate between the shell's inner core and the turning
surface, since as D modes they satisfy $|\tilde\Omega_p(\eta R)| <
2\Omega(\eta R)$. In contrast, the DT modes above the upper solid
curve, and those below the lower solid curve, propagate between the
turning surface and the shell's surface. Interestingly, modes with
$|m| > 2$ may exhibit two turning surfaces that are located between
the shell's inner core and the shell's surface. This brings the
possibility of waves naturally bouncing between two turning surfaces
without interacting with the shell's boundaries. Such modes are
depicted by grey areas in the bottom-right panel in
Fig.~\ref{turning}.  Note, however, that this particular set of
non-axisymmetric modes will necessarily feature a critical layer
(corotation resonance) between both turning surfaces, where
$\tilde\Omega_p$ vanishes.  {\color{black}The location of such critical
  layers is given by $r=R(-\Omega_p / m\Omega_{\rm ref})^{1/\sigma}$.
  For $m > 0$, modes with $\sigma \leq 0$ have a critical layer if $-m
  \eta^{\sigma} \leq \Omega_p / \Omega_{\rm ref} \leq -m$, and those
  with $\sigma \geq 0$ if $-m \leq \Omega_p / \Omega_{\rm ref}\leq -m
  \eta^{\sigma}$.  Modes that exhibit a critical layer within the
  shell are located between the two orange solid curves in
  Fig.~\ref{turning}. They will be further discussed in
  Sections~\ref{sec:inv_corot_shell}
  and~{\ref{sec:num_nonaxi_shell}}.}

The range of eigenfrequencies $\Omega_p$ for which inertial modes of
oscillation may {\color{black}exist} in a spherical shell with
$\eta=0.35$ is displayed in Fig.~\ref{turning2} for $m=0$, 1, 2, and
4.  Compared to solid-body rotation, inertial modes with shellular
rotation may {\color{black}exist} at much larger frequencies,
especially if the rotation profile is a decreasing function of radius.

\begin{figure}
  \centering
  \resizebox{\hsize}{!}
   {
    \includegraphics{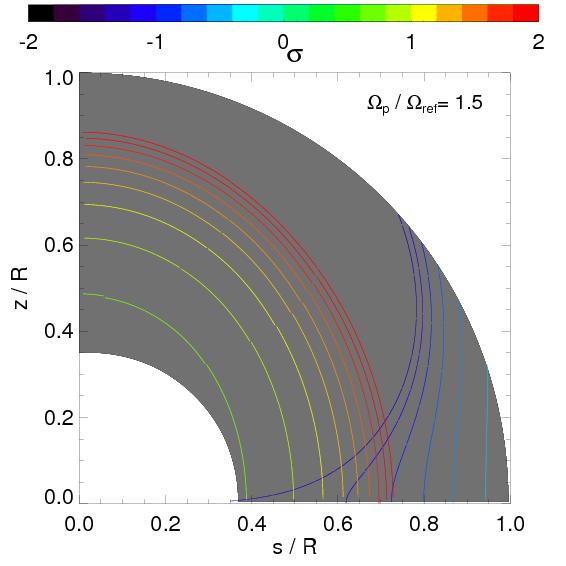}
    \includegraphics{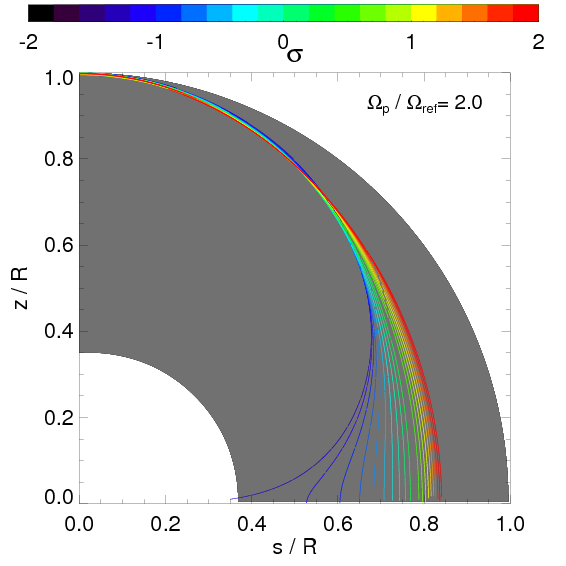}
    \includegraphics{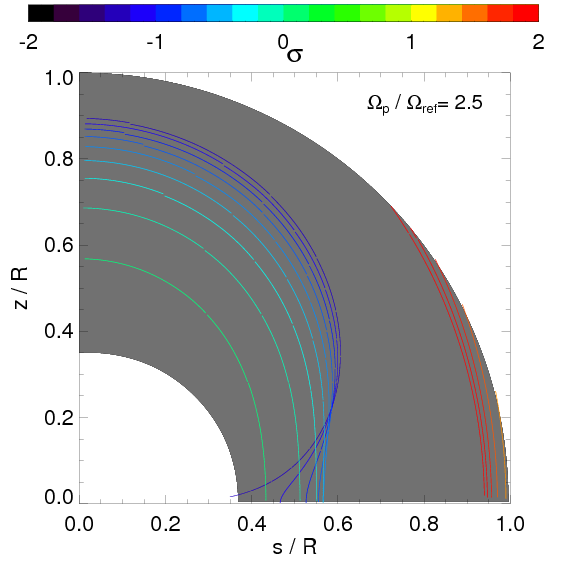}
  } 
  \caption{\label{loc_turning}Location of turning surfaces for $m=0$
    modes with varying the power-law index of the shellular rotation
    profile, for $\Omega_p = 1.5 \Omega_{\rm ref}$ (left panel),
    $\Omega_p = 2\Omega_{\rm ref}$ (middle panel), and $\Omega_p =
    2.5\Omega_{\rm ref}$ (right panel).}
\end{figure}
Finally, the location of turning surfaces, given by
Eq.~(\ref{eq:turning}), is depicted in a meridional slice in
Fig.~\ref{loc_turning} for $m=0$ modes with $\sigma$ ranging from -2
to 2, and increasing by 0.2. Results are shown for $\Omega_p /
\Omega_{\rm ref} = $ 1.5, 2, and 2.5 from left to right. For $\Omega_p
= 2\Omega_{\rm ref}$, turning surfaces exist for all values of
$\sigma$ but for $\sigma=0$ (corresponding to solid-body rotation;
note however that turning surfaces exist in the limit $|\sigma|
\rightarrow 0$).  In all three panels, DT modes with $\sigma < 0$
exist between the shell's inner core and the turning surface (they may
therefore {\color{black}exist} without reflexions on the shell's
surface).  Similarly, DT modes with $\sigma > 0$ exist between the
turning surface and the shell's surface, and may therefore
{\color{black}exist} without reflexions on the shell's inner core. This
case will be illustrated in Section~{\ref{sec:num_axi_shell}}, where
we will show that DT modes may feature shear layers following
attractors of characteristics in the absence of an inner core (the
necessary reflexion being provided by the shell's surface and the
turning surface).

\subsubsection{Critical latitudes}
Rays tangent to the shell's inner {\color{black}core} define a critical
latitude $\theta_{\rm i}$. They satisfy $dz/ds = -s/z$ with $\eta R =
\sqrt{s^2 + z^2}$. Combining Eqs.~(\ref{caract}) to~(\ref{Ax}), rays
intersect the shell's inner core at $s=s_{\rm c}$, with
\begin{equation}
	s_{\rm c} = \eta R \left( 1 - \frac{\tilde\Omega_p^2(\eta R)}{4\Omega^2(\eta R)} \right)^{1/2}.
	\label{critlat}
\end{equation}
A critical latitude $\theta_{\rm i}$ at the shell's inner core 
therefore exists for modes with $|\tilde\Omega_p(\eta R)| \leq 2
\Omega(\eta R)$, and verifies
\begin{equation}
	\sin\theta_{\rm i} = \frac{\tilde\Omega_p(\eta R)}{2\Omega(\eta R)},
	\label{critlat2}
\end{equation}
which is a straightforward generalization of the expression for
solid-body rotation. For fixed values of $\eta$, $m$ and $\Omega_p$,
and for $\sigma > 0$, $\sin\theta_{\rm i} \rightarrow 1$ as $\sigma$
increases. For $\sigma < 0$, $\sin\theta_{\rm i} \rightarrow 0$ as
$|\sigma|$ increases.

Similarly, it is easy to show that a critical latitude $\theta_{\rm
  o}$ exists at the shell's surface for those modes that satisfy
$|\tilde\Omega_p(R)| \leq 2 \Omega(R)$, with expression given by
\begin{equation}
  \sin\theta_{\rm o} = \frac{\tilde\Omega_p(R)}{2\Omega(R)},
	\label{critlat3}
\end{equation}
independently of $\sigma$. The outer critical latitude has 
therefore the same value as in the solid-body rotation case.

Critical latitudes also exist with cylindrical rotation. For our
cylindrical rotation profile, an inner critical latitude exists if
$\xi(s)$, given by Eq.~(\ref{cyl_xi}), is positive for $s \in [0,\eta
R]$. Similarly, an outer critical latitude exists if $\xi \geq 0
\;\forall s \in [0,R]$ (that is, for D modes only). There is, however,
no simple expressions for the inner and outer critical latitudes with
our cylindrical rotation profile.

\subsubsection{Dispersion relation, phase and group velocities}
\label{sec:inv_corot_shell}
{\color{black} 
In the short-wavelength approximation, the wave dispersion relation given by 
Eq.~(\ref{disp_app}) becomes}
\begin{equation}
  \tilde\Omega_p^2 (r) = 4\Omega^2(r) \frac{k_z^2}{ \|\vect{k}\|^2 } \times
  \left[
    1 + \frac{\sigma}{2} \frac{s^2}{r^2} \left(1 - \frac{zk_s}{sk_z}\right) 
  \right]
\label{disp1}
\end{equation}
{\color{black}for our shellular rotation profile}, with  
$\|\vect{k}\| = \sqrt{k_s^2 + k_z^2}$. {\color{black}We see that 
  the bracket term in the above equation can be negative: inertial 
  waves may become unstable for shellular rotation. As discussed 
  in the appendix, this is a form of the Goldreich-Schubert-Fricke instability 
  in the limit of an inviscid incompressible fluid \citep{gs67, fricke68}. 
  In that limit, Eq.~(\ref{disp_app}) indicates that the instability can in principle 
  occur whenever $\partial\Omega / \partial z \neq 0$ (or, equivalently, 
  when $A_z \neq 0$). Assuming that inertial waves remain stable 
  against our shellular rotation profile,} we introduce
\begin{equation} {\cal B} = \left[ 1 + \frac{\sigma}{2}
    \frac{s^2}{r^2} \left(1 - \frac{zk_s}{sk_z}\right) \right]^{1/2},
\end{equation}
and Eq.~(\ref{disp1}) reduces to $\tilde\Omega_p(r) = \pm 2 {\cal B}
\Omega(r) |k_z| / \|\vect{k}\|$. Note that ${\cal B} = 1$ for
solid-body rotation. {\color{black}In the rotating frame,} the phase
velocity reads
\begin{equation}
	\vect{v_p} = \pm 2 {\cal B} \Omega(r) \frac{k_z \vect{k}}{\|\vect{k}\|^3},
	\label{vphase}
\end{equation}
and the group velocity is given by
\begin{equation}
	\vect{v_g} = \pm 2 \Omega(r) \frac{k_s}{\|\vect{k}\|^3} (-k_z \vect{e_s} + k_s \vect{e_z}) \times 
	\left[ {\cal B} + \frac{\sigma}{4}\frac{sz}{r^2} \frac{\|\vect{k}\|^2}{k_s k_z} {\cal B}^{-1} \right].
	\label{vgroup}
\end{equation}

We now examine the case where a critical layer forms inside the shell,
where $\tilde\Omega_p (r) = 0$. Inspection of Eq.~(\ref{disp1}) shows
that there are formally three cases where $\tilde\Omega_p (r)$ may
vanish:
\begin{itemize}
\item $k_s \rightarrow \infty$ at finite $k_z$. In this case,
  $\vect{v_p} = \vect{0}$ and $\vect{v_g} = \vect{0}$ at corotation,
  which is similar to the behavior obtained with cylindrical rotation
  (inertial waves do not cross corotation {\color{black}in such case}).
\item $k_z \rightarrow 0$ at finite $k_s$, in which case $\vect{v_p} =
  \vect{0}$, $\vect{v_g} \cdot \vect{e_s} = 0$, and $|\vect{v_g} \cdot
  \vect{e_z}| \rightarrow \infty$. Surprisingly, this case indicates
  that inertial waves may propagate across a critical layer (with
  locally vertical paths of characteristics).
\item ${\cal B} \rightarrow 0$, which implies that $\vect{v_p} =
  \vect{0}$, $|\vect{v_g} \cdot \vect{e_s}| \rightarrow \infty$ and
  $|\vect{v_g} \cdot \vect{e_z}| \rightarrow \infty$. This case
  {\color{black}also} formally shows that inertial waves may cross a critical
  layer. The condition ${\cal B} = 0$ can be recast as
\begin{equation}
  {\frac{dz}{ds}\Big|}_{\rm crit} \equiv -{\frac{k_s}{k_z}\Big|}_{\tilde\Omega_p(r) = 0} = -\frac{s}{z} \left( 1 + \frac{2}{\sigma}\frac{r^2}{s^2} \right),
  \label{slopecrit}
\end{equation}
which shows that, {\color{black}if ${\cal B} = 0$ at corotation, then}
paths of characteristics have a finite slope upon crossing 
the critical layer.
\end{itemize}

The propagation properties of inertial waves in viscous fluids with a
critical layer will be examined in Section~\ref{sec:num_nonaxi_shell}
with the results of numerical calculations. We stress again that the
fact the phase speed vanishes at corotation indicates that non-linear
effects should occur near this location, which are not captured by the
present linear analysis, nor by the numerical solutions of
Section~\ref{sec:num_nonaxi_shell}.

\section{Viscous problem: shear layers, behaviour at corotation resonances, 
comparison to inviscid analysis}
\label{sec:shear}
We have described in Section~\ref{sec:inv} the propagation properties
of inertial waves in a spherical shell with (specific) cylindrical and
shellular rotation profiles. We have shown that in both cases, modes
of oscillations satisfy in the inviscid limit a second-order PDE of
mixed type. Two families of eigenmodes can therefore be distinguished:
D modes, which may propagate throughout the entire shell (the
hyperbolic domain covers the whole shell), and DT modes, which feature
turning surfaces inside the shell (the hyperbolic domain covers part
of the shell). Compared to solid-body rotation, the presence of
turning surfaces with differential rotation {\color{black} tends} to
broaden the range of frequencies at which inertial waves may propagate
in a spherical shell.

An interesting result from our analysis in Section~\ref{sec:inv} is
that paths of characteristics depend on the azimuthal wavenumber $m$
(through the space-varying Doppler-shifted frequency).  In contrast to
solid-body rotation, the structure of non-axisymmetric modes cannot be
simply inferred from that of axisymmetric modes.  However, as far as
their dynamical properties are concerned, $m \neq 0$ and $m=0$ modes
only differ by the possible presence of corotation resonances for $m
\neq 0$. We will therefore examine the general properties of viscous
modes of oscillations for $m=0$ modes only. This is done in
Section~{\ref{sec:num_axi}}, where we take the same rotation profiles
as in Section~{\ref{sec:inv}}.  We show that the structure of shear
layers in the viscous problem closely follows the paths of
characteristics derived in the inviscid problem.  We particularly
focus on {\color{black}singular modes featuring a shear layer that
  converges towards a short-period attractor. Some of these singular
  modes form through reflexions at turning surfaces and may therefore
  exist in the absence of an inner core. A qualitative analysis of the
  occurrence of modes with quasi-periodic orbits of characteristics is
  also done at Section~\ref{sec:quasiper}.}  We finally
{\color{black}examine} in Section~{\ref{sec:num_nonaxi}} a few $m \neq
0$ viscous eigenmodes with corotation resonances.

\subsection{Numerical method}
\label{sec:num}
Our numerical simulations solve the linearized differential system
\eq{eq:vort} and related stress-free boundary conditions, using a
spectral approach. Fields are decomposed into spherical harmonics
\cite[see, e.g.,][]{Rieutord87}, with the perturbed velocity expanded
as

\begin{equation}
  \vect{u}(r,\theta,\varphi) = \sum_{\ell=0}^{\infty} \sum_{m=-\ell}^{\ell} 
  u_m^{\ell}(r) \vect{R_{\ell}^m} 
  + v_m^{\ell}(r) \vect{S_{\ell}^m} 
  + w_m^{\ell}(r) \vect{T_{\ell}^m},
\end{equation}
with $\vect{R_{\ell}^m} = Y_{\ell}^m(\theta,\varphi) \vect{e_r}$,
$\vect{S_{\ell}^m} = \vect{\nabla} Y_{\ell}^m$, and $\vect{T_{\ell}^m}
= \vect{\nabla} \times \vect{R_{\ell}^m}$, and where $Y_{\ell}^m$
denote the usual spherical harmonics normalized on the sphere of unit
radius. In the radial direction, equations are discretized on
Gauss-Lobatto collocation nodes associated with Chebyshev
polynomials. Equations are truncated at order $L$ for the spherical
harmonics basis, and at order $N_r$ for the Chebyshev basis. The
incomplete Arnoldi-Chebyshev algorithm is used to compute pairs of
complex eigenvalues ($\Omega_p$) and eigenvectors, given an initial
guess value for $\Omega_p$ \cite[for details on this numerical method,
see][]{VRBF07}. Values of $N_r$ and $L$ are mode-dependent, and will
be specified below. {\color{black}Unless otherwise stated, all modes
  are calculated assuming symmetry with respect to the equatorial
  plane}.

\subsubsection{Shellular rotation}
For the shellular rotation profile that we consider, $\Omega(r) /
\Omega_{\rm ref}= (r/R)^{\sigma}$, projecting Eqs.~\eq{eq:vort} onto
$\vect{R_{\ell}^m}$ and $\vect{T_{\ell}^m}$ gives the following two
equations:

\begin{eqnarray}
\left.\begin{aligned}
%
\lefteqn{
E \Delta_{\ell} w_m^{\ell} 
+ \left[  
\frac{2im\Omega(r)}{\ell(\ell +1)} 
- i\tilde{\Omega}_p(r)
\right]w_m^{\ell}  = 
}\\
&&   
-2\Omega(r) \left[
A(\ell,m) r^{\ell -1} \frac{d(r^{2-\ell}u_m^{\ell -1})}{dr}
+ A(\ell+1,m) r^{-\ell -2} \frac{d(r^{\ell +3}u_m^{\ell +1})}{dr}
\right]
\\
&&
+r \frac{d\Omega}{dr} 
\left[ -(\ell +1)A(\ell+1,m) u_m^{\ell +1} + \ell A(\ell,m) u_m^{\ell -1} \right],
\\
\\
\lefteqn{
E \Delta_{\ell}\Delta_{\ell} (r u_m^{\ell})
+
\left[
\frac{2im\Omega(r)}{\ell(\ell +1)} 
- i\tilde{\Omega}_p(r)
\right]\Delta_{\ell} (ru_m^{\ell})
 = 
} \\
&&
2\Omega(r)\left[ 
B(\ell,m) r^{\ell -1} \frac{d(r^{1-\ell}w_m^{\ell -1})}{dr} 
+ B(\ell+1,m) r^{-\ell -2} \frac{d(r^{\ell +2}w_m^{\ell +1})}{dr}
\right]
\\
&&
- imr \frac{d^2 \Omega}{dr^2} u_m^{\ell}
- 2im\frac{d\Omega}{dr} (u_m^{\ell} + v_m^{\ell})
\\
&&
+ 2\frac{d\Omega}{dr} \left[
 B(\ell,m) w_m^{\ell -1} 
 + B(\ell+1,m) w_m^{\ell +1}
\right],
\end{aligned}
 \;\;\;\right\}
 \qquad \text{~}
\label{eq_shell}
\end{eqnarray}
where
\[
A(\ell,m) = \frac{1}{\ell^2}\left( \frac{{\ell}^2 - m^2}{4{\ell}^2 - 1} \right)^{1/2}, 
\;\;B(\ell,m) = \ell^2 (\ell^2-1) A(\ell,m),
\;\;\Delta_{\ell} = \frac{1}{r}\frac{d^2}{dr^2}r - \frac{\ell(\ell+1)}{r^2}.
\]
In Eqs.~(\ref{eq_shell}), since the fluid is incompressible,
\begin{equation}
v_m^{\ell} = \frac{1}{\ell(\ell +1)r} \frac{d(r^2 u_m^{\ell})}{dr}.
\end{equation}
For solid-body rotation, $\Omega(r)=\Omega_{\rm ref}$ and
Eqs.~(\ref{eq_shell}) reduce to equations (2.2) of \cite{RV97} (note
that these authors defined the Ekman number as $E=\nu / 2R^2
\Omega_{\rm ref}$, while our definition does not include the factor
2).  After projection on the spherical harmonics, stress-free boundary
conditions read
\begin{equation}
\ulm = \frac{d^2(r\ulm)}{dr^2} = \frac{d}{dr}\left( \frac{{\color{black}\wlm}}{r} \right) = 0
\label{BC_shell}
\end{equation}
for the radial functions taken at $r=\eta R$ or $r=R$.
Eqs.~(\ref{eq_shell}) to~(\ref{BC_shell}) can be recast as an
eigenvalue problem featuring a tridiagonal block matrix,
{\color{black}like for solid-body rotation}. More details into the
numerical scheme can be found in \cite{RV97}.

\subsubsection{Cylindrical rotation}
For our cylindrical rotation profile, $\Omega(s) / \Omega_{\rm ref}= 1
+ \varepsilon (s/R)^2$, projecting Eqs.~\eq{eq:vort} onto
$\vect{R_{\ell}^m}$ and $\vect{T_{\ell}^m}$ gives, after some tedious
algebra, two long differential equations, which we do not write for
the sake of legibility. The second-order differential equation
satisfied by $w_m^{\ell}$ can be cast as a linear combination of
$u_m^{\ell \pm 3}$, $du_m^{\ell \pm 3}/dr$, $u_m^{\ell \pm 1}$,
$du_m^{\ell \pm 1}/dr$ and $w_m^{\ell \pm 2}$ (the term in front of
$w_m^{\ell \pm 2}$ vanishes for $m=0$). The fourth-order differential
equation for $u_m^{\ell}$ exhibits terms in $w_m^{\ell \pm 3}$,
$dw_m^{\ell \pm 3}/dr$, $w_m^{\ell \pm 1}$, $dw_m^{\ell \pm 1}/dr$,
$u_m^{\ell \pm 2}$, $du_m^{\ell \pm 2}/dr$ and $d^2u_m^{\ell \pm
  2}/dr^2$ (the terms in front of $u_m^{\ell \pm 2}$, $du_m^{\ell \pm
  2}/dr$ and $d^2u_m^{\ell \pm 2}/dr^2$ vanish for $m=0$).  Along with
stress-free boundary conditions, these equations take the form of an
eigenvalue problem where the matrix to invert has blocks composed of
(up to) seven bands. The coupling between harmonics of ranks $\ell$,
$\ell \pm 1$, $\ell \pm 2$ and $\ell \pm 3$ arises from our particular
cylindrical profile.

\subsection{Axisymmetric eigenmodes: a few illustrative cases}
\label{sec:num_axi}
We present in this section the results of simulations for axisymmetric
eigenmodes with moderate viscosity, the default Ekman number being $E
= 10^{-8}$. Unless otherwise stated, our simulations are carried out
with {\color{black}a shell's aspect ratio} $\eta=0.35$. The propagation
properties of shear layers for weakly-damped eigenmodes is compared
with the analytic description based on paths of characteristics,
detailed in Section~\ref{sec:inv}. Cylindrical and shellular rotation
profiles are examined in Sections~\ref{sec:num_axi_cyl}
and~\ref{sec:num_axi_shell}, respectively. {\color{black}A qualitative
  analysis of the occurrence of modes with quasi-periodic orbits of
  characteristics follows in Section~\ref{sec:quasiper}}.

\subsubsection{Cylindrical rotation}
\label{sec:num_axi_cyl}
We describe in this paragraph three representative $m=0$ eigenmodes
obtained with our cylindrical rotation profile: a D mode (the
hyperbolic domain covers the whole shell), and two DT modes (the
hyperbolic domain covers a fraction of the shell). All three
calculations have a spectral resolution $N_r = 450 \times L = 800$.

\begin{figure}
  \centering
  \resizebox{\hsize}{!}
  {
    \includegraphics[width=0.5\hsize]{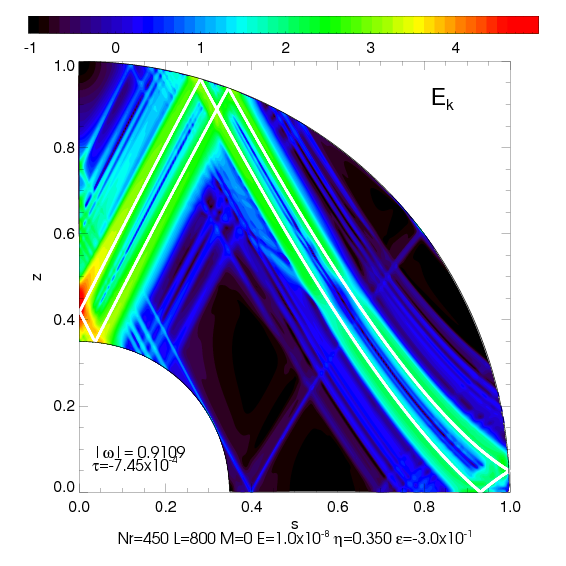}
    \includegraphics[width=0.5\hsize]{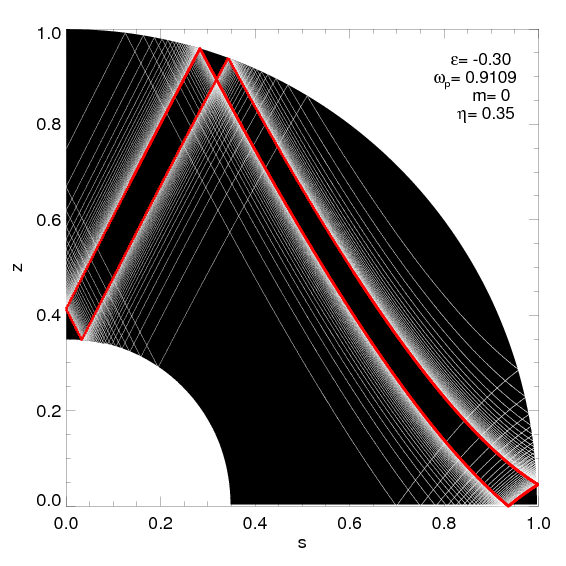}
  }
   \resizebox{\hsize}{!}
  {
    \includegraphics[width=\hsize]{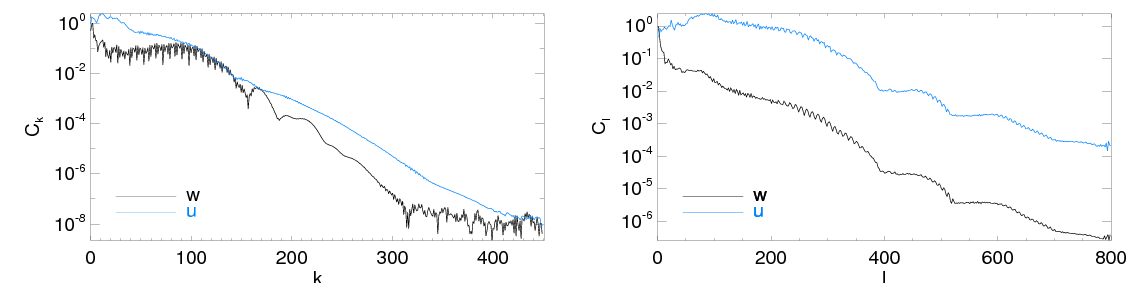}
  }
  \caption{\label{nice_cyl}Top left: meridional cut of the kinetic
    energy for a D mode with eigenfrequency $\Omega_p \approx 0.91
    \Omega_{\rm ref}$, obtained with $E=10^{-8}$ and cylindrical
    rotation profile $\Omega(s) / \Omega_{\rm ref} = 1 - 0.3
    (s/R)^2$. An attractor of characteristics is overplotted by a
    solid white curve.  Top right: a few paths of characteristics are
    shown to converge towards the aforementioned attractor, which is
    displayed here by a red curve. Bottom: spectral content of the
    radial ($u$) and orthoradial ($w$) components of the velocity
    field for this mode.  Chebyshev and spherical harmonic
    coefficients are shown in the left and right panels,
    respectively.}
\end{figure}
Fig.~\ref{nice_cyl} displays the results for a D mode with
eigenfrequency $\Omega_p \approx 0.91\Omega_{\rm ref}$, obtained for
$\varepsilon = -0.3$ ({\color{black}recall that for cylindrical
  rotation, $\Omega_{\rm ref}$ denotes the shell's angular frequency
  at the rotation axis}).  The top-left panel depicts contours of the
mode's kinetic energy in a meridional
{\color{black}quarter-plane\footnote{In all our contour plots, the
    simulation parameters are summarized below the x-axis (spectral
    resolution $N_r \times L$, azimuthal wavenumber $m$, Ekman number
    $E$, shell's aspect ratio $\eta$, differential rotation rate
    $\varepsilon$ or $\sigma$ for cylindrical or shellular rotation
    profiles, respectively). Similarly, the complex eigenfrequency is
    indicated in the bottom-left corner of the shell: the absolute
    value of its real part is denoted by $|\omega|$ ({\color{black} in
      units of $\Omega_{\rm ref}$)}, and its imaginary part by $\tau$
    ({\color{black} in units of $\Omega^{-1}_{\rm ref}$}; negative
    values of $\tau$ correspond to damped modes).}}.  Results are
shown only for positive values of $z$, for reasons of symmetry exposed
in Section~{\ref{sec:inv_cyl}}.  This D mode essentially
{\color{black}features} a shear layer {\color{black}following} a
short-period {\color{black}wave attractor, which is} overplotted by a
thick white curve. The mode's kinetic energy is maximum near the
rotation axis, like in the case of solid-body rotation \citep{RV97}.
Note also the presence of another shear layer tangent to the shell's
inner core. {\color{black}As already highlighted in
  Section~{\ref{sec:inv_cyl}}, rays in a meridional plane are curved
  by differential rotation}. This is also illustrated in the top-right
panel of Fig.~\ref{nice_cyl}, which displays the paths of
characteristics obtained by solving Eq.~(\ref{cyl_caract}), given a
particular initial location (here at $s=0$ and $z = 0.75 R$). The
patterns of the viscous shear layer (left panel) and of the paths of
characteristics under the short-wavelength approximation (right
panel), are in excellent agreement. Paths of characteristics focus
towards the aforementioned wave attractor, depicted by a red curve in
this panel. Upon varying initial conditions, we have checked that this
is the only attractor in the shell for this particular
eigenfrequency. The bottom panel of Fig.~\ref{nice_cyl} shows the good
convergence in spectral resolution for this mode.

\begin{figure}
  \centering
  \resizebox{\hsize}{!}
  {
    \includegraphics[width=0.5\hsize]{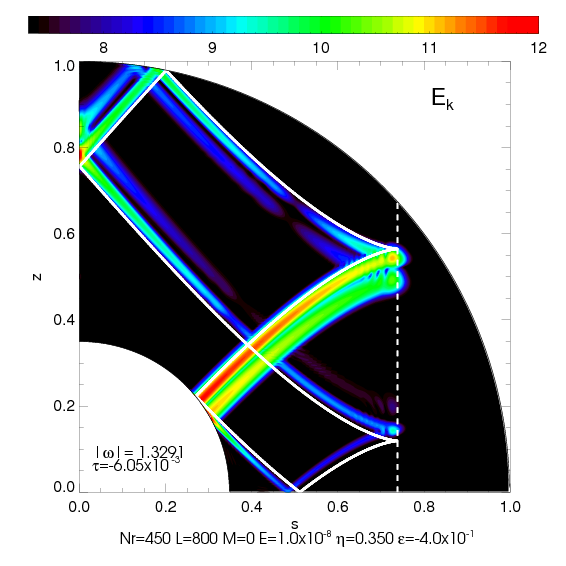}
    \includegraphics[width=0.5\hsize]{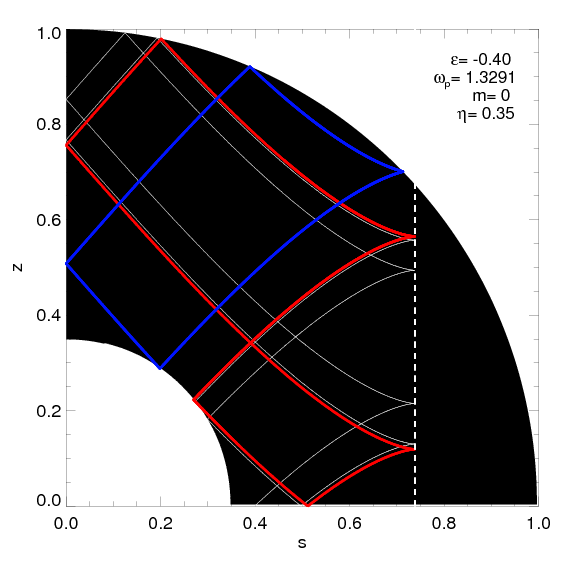}
  }
  \caption{\label{nice2_cyl}Left: meridional slice of the kinetic
    energy for a DT mode with $\Omega_p \approx 1.33 \Omega_{\rm
      ref}$, $E=10^{-8}$, and $\varepsilon = -0.4$. The shear layer
    follows a wave attractor (solid curve). The dashed line shows the
    location of the turning surface inferred from
    Eq.~(\ref{cyl_xi}). Right: a few paths of characteristics are
    depicted by white curves, which converge toward the aforementioned
    wave attractor (red curve).  A second wave attractor is shown in
    blue.}
\end{figure}

We present in Fig.~\ref{nice2_cyl} the results for an $m=0$ DT mode
with $\Omega_p \approx 1.33 \Omega_{\rm ref}$ and $\varepsilon =
-0.4$.  The left panel shows a meridional cut of the kinetic energy
for this mode, which exhibits again a shear layer following an
attractor of characteristics. The mode's pattern features reflexions
on the inner and outer edges of the shell, as well as on a turning
surface located at $s \sim 0.75 R$. The radius at which the shear
layer bounces off is in good agreement with the (numerical) solution
to $\xi(s) = 0$, shown by a dashed line. A few paths of
characteristics, obtained again by solving Eq.~(\ref{cyl_caract}), are
shown by white curves in the right panel of Fig.~\ref{nice2_cyl},
which rapidly converge toward the above attractor (in red in this
panel). We point out that there exists another attractor for
{\color{black} this} eigenfrequency and {\color{black} this}
differential rotation rate, which is overplotted by a blue curve.

\begin{figure}
  \centering \resizebox{0.5\hsize}{!}  {
   \includegraphics{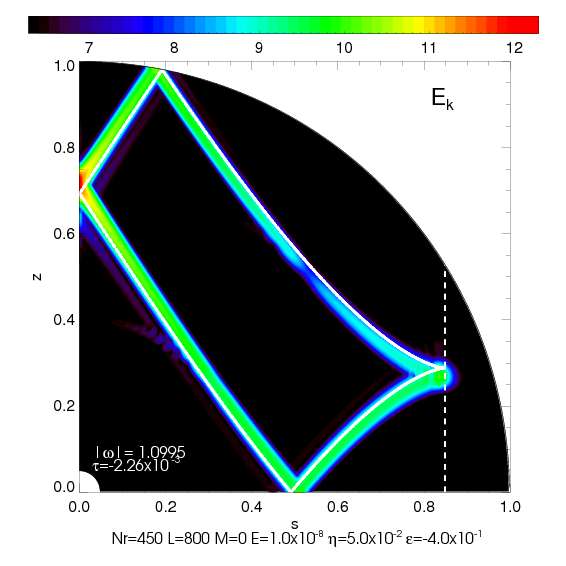}
  } 
  \caption{\label{nice3_cyl}Meridional slice of the kinetic energy for
    a DT mode with $\Omega_p \approx 1.10 \Omega_{\rm ref}$,
    $E=10^{-8}$, $\eta=0.05$ and $\varepsilon = -0.4$. The turning
    surface is shown by a dashed line. For an arbitrarily small inner
    core, the shear layer follows an attractor of characteristics,
    shown by the solid curve.}
\end{figure}

An interesting DT mode is finally illustrated in Fig.~\ref{nice3_cyl}
for $\Omega_p \approx 1.10 \Omega_{\rm ref}$ and $\varepsilon = -0.4$,
where a shear layer follows an attractor for an arbitrarily small
inner core.  In this particular calculation, $\eta=0.05$.  This
example highlights that, in contrast to solid-body rotation,
attractors may be formed by reflexions on one of the shell's edges
(the surface in this case) and a turning surface.  This property is
also shared by gravito-inertial modes in stably-stratified rotating
fluids \citep{Fried82, Dintrans99}.  The mode shown in
Fig.~\ref{nice3_cyl} indicates that, with cylindrical rotation, wave
attractors may exist independently of a putative inner core inside the
shell. We show below that the same is true with shellular rotation.

\subsubsection{Shellular rotation}
\label{sec:num_axi_shell}
\begin{figure}
  \centering
  \resizebox{\hsize}{!}
  {
   \includegraphics{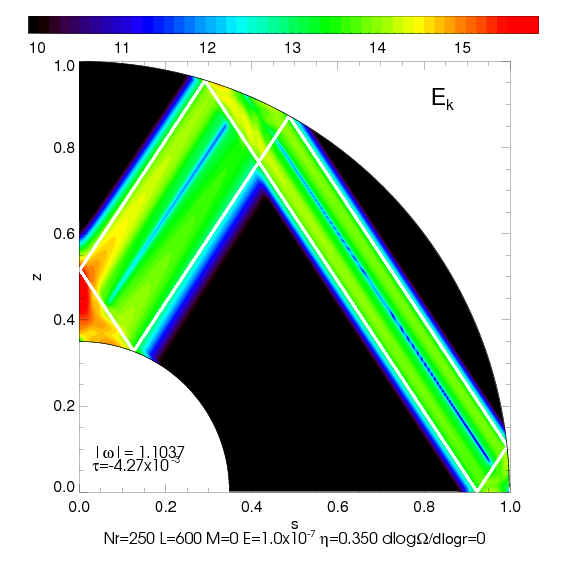}
   \includegraphics{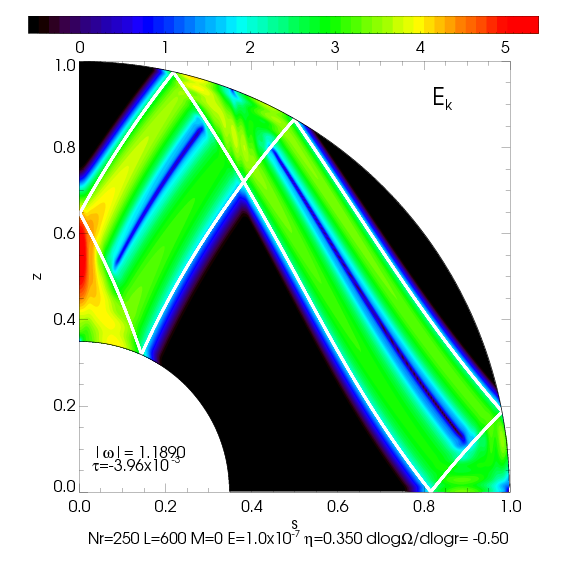}
   }
   \resizebox{\hsize}{!}
   {
   \includegraphics{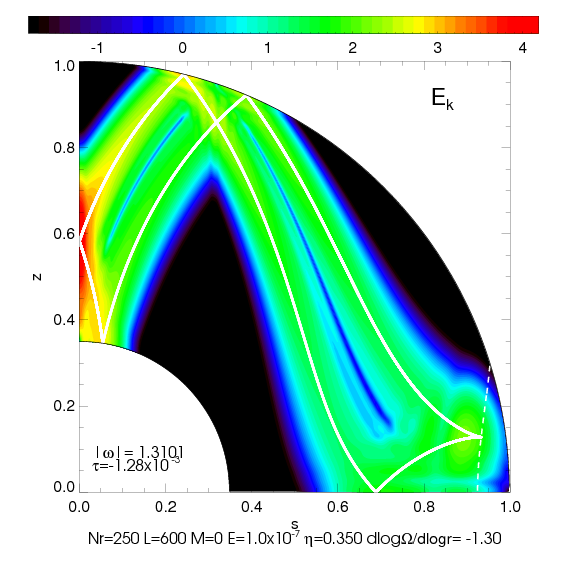}
   \includegraphics{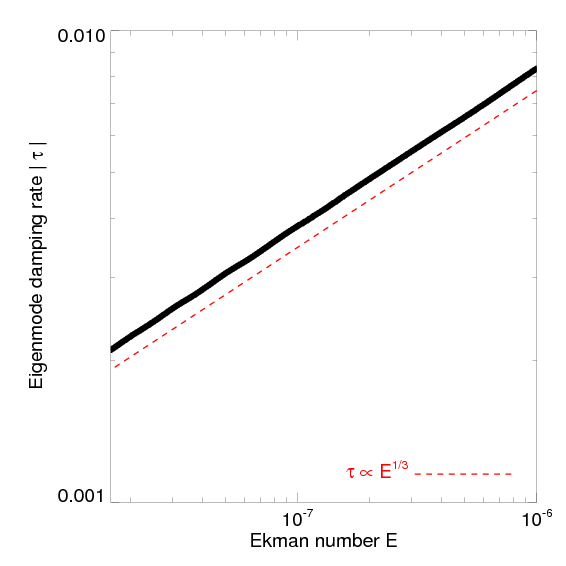}
   }
   \caption{\label{givenmode_omslope}Meridional cuts of kinetic energy
     for a given eigenmode with $E=10^{-7}$ and $\sigma=0$ (top-left),
     $-0.5$ (top-right), and $-1.3$ (bottom-left). The shear layer for
     this mode follows an attractor overplotted by a solid curve in
     each of these panels.  The dashed curve in the bottom-left panel
     displays the location of a turning surface, given by
     Eq.~(\ref{eq:turning}). {\color{black}The bottom-right panel shows
       that the damping rate of previous mode with $\sigma=-1$ scales
       as $E^{1/3}$, with $E$ the Ekman number}.}
\end{figure}
\begin{figure}
  \centering
  \resizebox{\hsize}{!}
  {
    \includegraphics[width=0.5\hsize]{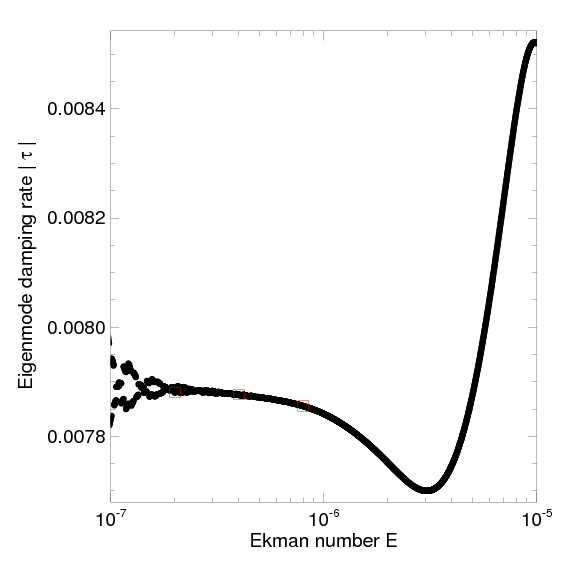}
    \includegraphics[width=0.5\hsize]{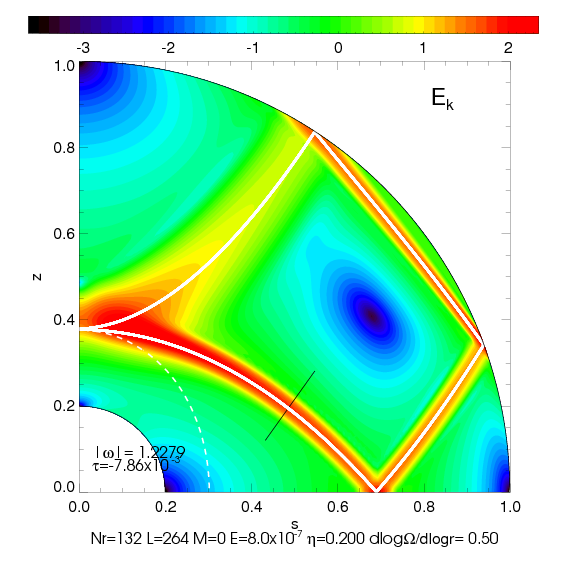}
  }
   \resizebox{0.7\hsize}{!}
  {
    \includegraphics{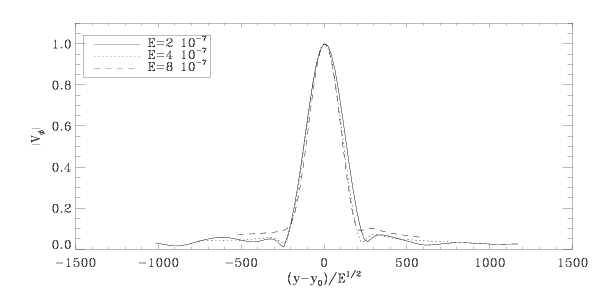}
  }
  \caption{\label{quasi}Top-left: evolution of the damping rate
    ($|\tau|$) with decreasing Ekman number for a DT eigenmode with
    shellular rotation, $\Omega(r) \propto r^{0.5}$, $\eta=0.2$,
    {\color{black} which features a detached shear layer tangent to a
      turning surface at the location of the rotation axis}.
    Top-right: contours of kinetic energy for the mode depicted by the
    rightmost {\color{black} open} square in the top-left panel ($E
    \approx 8.0 \times 10^{-7}$). The thick white curve shows the
    attractor of characteristics followed by the shear layer, and the
    dashed curve the location of a turning surface that encompasses
    the inner core. Bottom: cut of the perturbed velocity across the
    shear layer (at the location depicted by a black line in the
    top-right panel) for the three modes displayed with {\color{black}
      open} square symbols in the top-left panel.}
\end{figure}
We describe in this paragraph a few axisymmetric ($m=0$) eigenmodes
obtained with our shellular rotation profile, $\Omega(r) / \Omega_{\rm
  ref}= (r/R)^{\sigma}$ {\color{black}(here, $\Omega_{\rm ref}$ denotes
  the angular frequency at the shell's surface)}.  An overview of the
impact of shellular rotation is illustrated in
Fig.~\ref{givenmode_omslope} for a singular mode with eigenfrequency
$\Omega_p \approx 1.11 \Omega_{\rm ref}$ for solid-body
rotation. Contours of the kinetic energy in a meridional
{\color{black}quarter-plane} are shown for $\sigma$ equal to 0
(top-left panel), $-0.5$ (top-right), and $-1.3$ (bottom-left).
Calculations were carried out as a sequence, the eigenfrequency of the
least-damped mode obtained with differential rotation rate $\sigma$
being used as the guess frequency for the next run with differential
rotation rate $\sigma - d\sigma$. In this series of runs, $d\sigma =
0.01$, $E=10^{-7}$, and the spectral resolution is $N_r=250 \times
L=600$.  The contour panels in Fig.~\ref{givenmode_omslope} show how
the shape of the shear layer, {\color{black}and thus that of the wave
  attractor it focuses on (which curve)}, is progressively deformed by
shellular rotation. Note the presence of a turning surface for $\sigma
= -1.3$ {\color{black}(dashed curve)}.  Not surprisingly,
{\color{black}the mode's eigenfrequency increases with increasing the
  steepness $\sigma$ of the rotation profile}.  We also observe that
the mode's damping rate decreases in absolute value for steeper
rotation profiles.  Some insight into the asymptotic behavior of this
mode at small Ekman numbers is given by the bottom-right panel in
Fig.~\ref{givenmode_omslope}.  The damping rate {\color{black}of the
  mode with $\sigma=-1$} is displayed for Ekman numbers varying from
$10^{-6}$ down to about $2\times 10^{-8}$. Runs were also performed as
a series, the eigenfrequency of the least-damped mode obtained at
Ekman number $E$ being used as the guess frequency for the run with
Ekman number $E-dE$ (the spectral resolution is progressively
increased with decreasing $E$). The dashed line shows that the damping
rate {\color{black}of this particular mode} scales proportional to
$E^{1/3}$, a scaling that is reminiscent of the solid-body rotation
case \citep{RV97}.

The aforementioned $E^{1/3}$ scaling of the mode's damping rate
{\color{black} is} not the only possible scaling for shellular rotation
in the limit of vanishing viscosities.  This is illustrated in the
top-left panel of Fig.~\ref{quasi} {\color{black} for a particular DT
  mode having a turning surface that encompasses the inner core, and
  for which the damping rate is found to be roughly independent of
  $E$.}  This property is reminiscent of quasi-modes associated with
critical layers \cite[see, e.g.,][]{ledizes04}. However, the physical
mechanism leading to such a damping rate must be different because the
DT mode we consider is axisymmetric, so that no corotation singularity
exists.  Contours of kinetic energy for this new type of quasi-modes
are shown in the top-right panel of Fig.~\ref{quasi}, where we see
that the mode is dominated by a shear layer following a wave attractor
{\color{black} tangent to the turning surface at the location of the
  rotation axis (note that, since the turning surface encompasses the
  inner core at $r=0.35R$, this mode exists} for arbitrarily small
inner cores, akin to the mode shown in Fig.~\ref{nice3_cyl} for
cylindrical calculation).  {\color{black} We have carried out
  additional calculations with a slightly different eigenfrequency at
  given Ekman number, such that the wave attractor no longer hits the
  tangent critical point at the rotation axis. We found that the
  damping rate is no longer independent of $E$ for the same range of
  Ekman numbers as in the top-left panel of Fig.~\ref{quasi}. It
  suggests that the tangent critical point plays a prominent role in
  getting a damping rate independent of $E$.  } Such {\color{black}
  scaling} demands a very thin detached shear layer, namely
\od{E^{1/2}}-wide. This is indeed confirmed by the velocity profiles
{\color{black} depicted} in the bottom panel of Fig.~\ref{quasi} at
three Ekman numbers (shown by red squares in the top-left panel of
{\color{black} Fig.~\ref{quasi}}).  Of course, this numerical result
does not prove that this vigorously damped mode survives at
vanishingly small viscosities, and that it is thus a new type of
quasi-modes. {\color{black} Still,} it calls for a detailed analysis of
this new type of {\color{black} detached} shear layers, which are as
thin as Ekman boundary layers.

\begin{figure}
  \centering
  \resizebox{\hsize}{!}
  {
    \includegraphics[width=0.5\hsize]{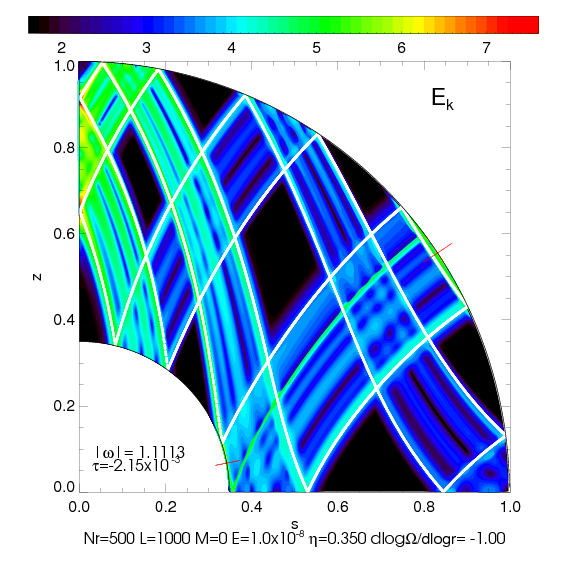}
    \includegraphics[width=0.5\hsize]{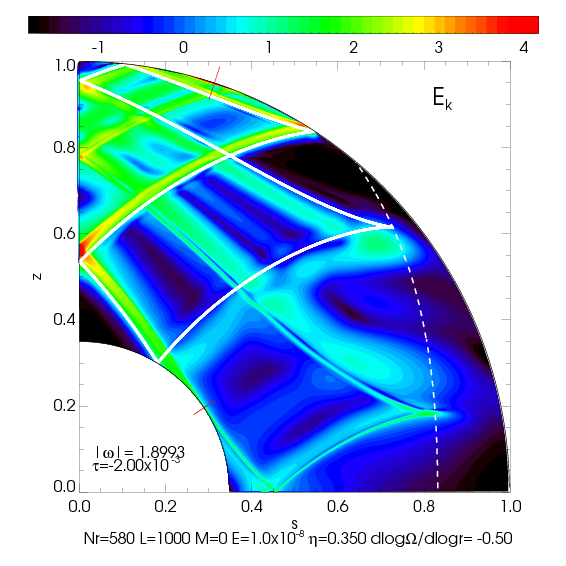}
  }
  \caption{\label{nice}Kinetic energy of two axisymmetric D and DT
    eigenmodes with shellular rotation, for $E=10^{-8}$. Left: D mode
    with $\Omega_p \approx 1.11 \Omega_{\rm ref}$ and $\Omega(r)
    \propto r^{-1}$. Right: DT mode with $\Omega_p \approx 1.90
    \Omega_{\rm ref}$ and $\Omega(r) \propto r^{-0.5}$. Red ticks show
    the values of the inner and outer critical latitudes given by
    Eqs.~(\ref{critlat2}) and~(\ref{critlat3}).}
\end{figure}
We now present the results of simulations for two D and DT eigenmodes
at smaller Ekman number ($E=10^{-8}$) and higher spectral resolution
($N_r=500$ or $N_r=580$, $L=1000$). Results are shown in
Fig.~\ref{nice}. A D mode with $\Omega_p \approx 1.11 \Omega_{\rm
  ref}$ and $\Omega \propto r^{-1}$ is shown in the left panel of this
figure. A shear layer is clearly visible that follows an attractor of
characteristics, calculated through Eq.~(\ref{caract}) and overplotted
by a white curve.  We note again the excellent agreement between the
overall pattern of the viscous shear layer, and the paths of
characteristics. Because of its finite width, the shear layer also
excites the critical latitude singularity, on both the inner and outer
edges of the shell.  The location of the inner and outer critical
latitudes, given by Eqs.~(\ref{critlat2}) and~(\ref{critlat3}), is
depicted by red ticks, and it is clear that the shear layer is tangent
to the inner and outer edges of the shell at these latitudes. Lastly,
the right panel of Fig.~\ref{nice} shows a DT mode with $\Omega_p
\approx 1.90 \Omega_{\rm ref}$ and $\Omega \propto r^{-0.5}$.  In this
case again, the mode's pattern suggests that part of the shear layer
approximately follows a wave attractor, shown by a white curve, and
that part of it is tangent to both the inner and outer edges of the
shell at the critical latitudes.

\subsubsection{Quasi-periodic orbits of characteristics}
\label{sec:quasiper}
{\color{black} The results presented in the previous paragraphs all
  feature modes that are associated with short-period attractors. These
modes are all singular at vanishing viscosity. One may however wonder
about the existence of regular modes, which exist in the inviscid case,
or at least modes that display shear layers only at unphysically small
viscosities. Such modes are necessarily associated with weak attractors
or with quasi-periodic orbits of characteristics.
To assess the occurrence of such orbits of characteristics,
  we have evaluated the Lyapunov exponent of various pairs
  of characteristics. The Lyapunov exponent measures the rate of
  convergence of two close characteristics after multiple reflections
  on the inner and outer edges of the shell, and on turning surfaces when
  present. For solid-body rotation, the latitudes of two successive
  reflections on the inner or outer edges can be related
  analytically, and the Lyapunov exponent can therefore be determined
  semi-analytically \citep{RGV01}. This is no longer the case with
  differential rotation: Lyapunov exponents must be evaluated
  numerically by integrating paths of characteristics given some
  initial conditions.

  For the symmetry reasons underlined in Sect.~\ref{sec:inv}, paths of
  characteristics are calculated in a meridional quarter-plane with
  reflections on to the rotation axis ($s=0$) and the equatorial axis
  ($z=0$). Because turning surfaces may prevent reflections on either 
  axis, we define the two following Lyapunov exponents associated with
  reflections on the equatorial axis (denoted by $\Lambda_s$) and on the 
  rotation axis ($\Lambda_z$):
\begin{equation}
  \Lambda_s = \lim_{N_s\rightarrow\infty}\frac{1}{N_s} \sum_{k=1}^{N_s} \ln \left| \frac{ds_{k+1}}{ds_{k}} \right|
  \;\;\;\;
  {\rm and}
  \;\;\;\;
  \Lambda_z =  \lim_{N_z\rightarrow\infty}\frac{1}{N_z} \sum_{k=1}^{N_z} \ln \left| \frac{dz_{k+1}}{dz_{k}} \right|,
\end{equation}
where $ds_k$ and $dz_k$ are the separation between two characteristics
after $k$ reflections on the equatorial and rotation axes,
respectively.

A negative Lyapunov exponent means that the two characteristics get
progressively closer to each other, converging towards an
attractor. The more negative this value, the faster the convergence
(and the stronger the attractor). In contrast, small-amplitude
Lyapunov exponents may correspond to quasi-periodic orbits of
characteristics. Because two attractors may co-exist in the shell at a
given eigenfrequency, and for averaging purposes, we have considered
five pairs of initial conditions and four regions of the shell located
near the rotation or equatorial axes, and near the shell's surface or
the inner core (that is, twenty pairs of characteristics).  The most
negative value amongst the (averaged) Lyapunov exponents obtained in
the above four regions defines what we now refer to {\it the} Lyapunov
exponent, and which we denote by $\Lambda$. If two attractors co-exist
in the shell, our method will thus retain the one with the shortest
period.  }

\begin{figure}
  \centering
  \resizebox{\hsize}{!}
   {
    \includegraphics{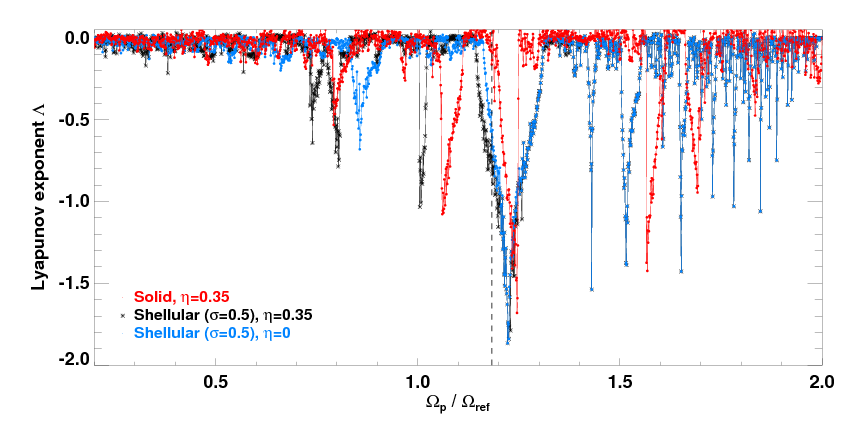}
   }
    \resizebox{\hsize}{!}
   {
    \includegraphics{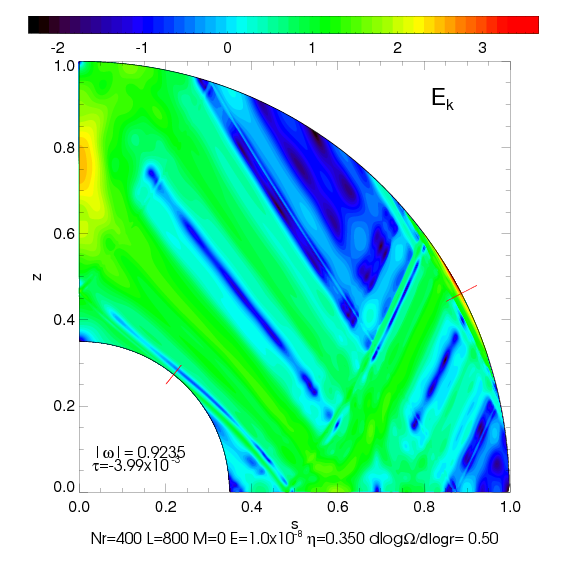}
    \includegraphics{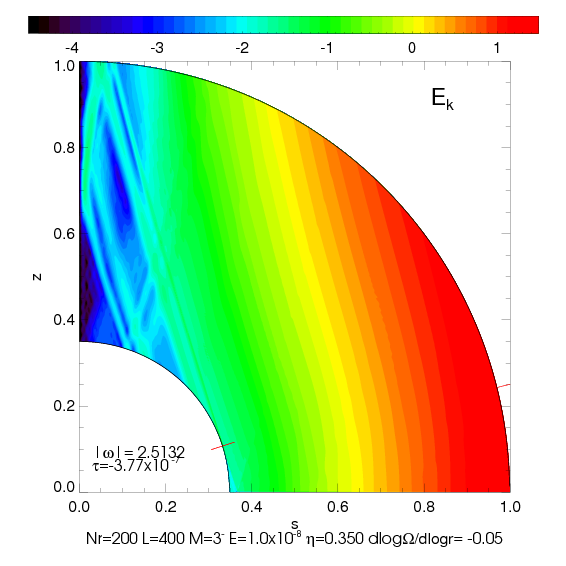}
  }
  \caption{\label{quasiper}{\color{black} Top: Lyapunov exponent
      obtained for $\eta=0.35$ and $m=0$.  Results with solid-body
      rotation (red curve) are compared to shellular rotation with
      $\sigma=0.5$ (black curve). The vertical dashed line displays
      the eigenfrequency that separates D and DT modes for this
      rotation profile. Results with shellular rotation, $\sigma=0.5$
      and for a full sphere ($\eta=0$) are overplotted in blue. Bottom
      left: meridional cut of the kinetic energy for an $m=0$ D mode
      with $\sigma = 0.5$ and $\Omega_p \approx 0.92 \Omega_{\rm
        ref}$, which features a weak attractor (the Lyapunov
      coefficient is a few percent). Bottom-right: meridional cut of
      the kinetic energy for an $m=3$ D mode with $\Omega_p \approx
      -2.5 \Omega_{\rm ref}$ and $\sigma = -0.05$. This mode has been
      obtained assuming asymmetry with respect to the equatorial
      plane.}}
\end{figure}
{\color{black} Lyapunov exponents obtained for $m=0$ and a few rotation
  profiles are displayed in the top panel of Fig.~\ref{quasiper}. The
  range of eigenfrequencies $[0.2 \Omega_{\rm ref}, 2 \Omega_{\rm
    ref}]$ is sampled with 1800 values ($\Omega_{\rm ref}$ is the
  angular frequency at the shell's surface).  The case with solid-body
  rotation and $\eta=0.35$ is depicted by a red curve. It can be
  compared with the semi-analytic results of \cite{RGV01}, see the
  left panel of their figure 6 (where $\Omega_p / 2\Omega_{\rm ref}$
  is displayed in x-axis). This comparison shows that our numerical
  method reproduces qualitatively well the ranges of frequencies at
  which strong attractors and quasi-periodic orbits of characteristics
  exist.  Quantitative differences inevitably exist, however, and are
  most significant for attractors with moderate Lyapunov exponents
  (that is, $\Lambda$ in the range $[-0.5,-0.2]$), which our method
  tends to underestimate.

  Results obtained with shellular rotation and $\sigma=0.5$ are
  overplotted by a black curve in the top panel of
  Fig.~\ref{quasiper}.  For this rotation profile, inertial modes
  exist for eigenfrequencies $|\Omega_p| \lesssim 2.2 \Omega_{\rm ref}
  $ (see, e.g., Fig.~\ref{turning}) and the transition from D to DT
  modes occurs at $|\Omega_p| \approx 1.18 \Omega_{\rm ref}$ (see
  Eq.~\ref{eq:dwmodes2}).  For $\Omega_p \leq 1.18 \Omega_{\rm ref}$
  (D modes only), the shellular case features the same attractors as
  in the case with solid-body rotation.  However, attractors are
  shifted toward lower frequencies with shellular rotation as the
  shell's interior rotates at a slower pace in this case. Lyapunov
  exponents also take very similar values since shellular rotation
  changes the overall shape of these attractors without altering their
  period.  For $\Omega_p \geq 1.18 \Omega_{\rm ref}$, the shellular
  case displays many more attractors, which form through reflections
  on to turning surfaces that encompass the inner core.  Finally,
  results obtained with the same shellular profile, but for the full
  sphere ($\eta=0$) are also shown by the blue curve. The full sphere
  only supports DT modes in this case. For $\Omega_p \gtrsim 1.4
  \Omega_{\rm ref}$, results are identical to those obtained with
  $\eta=0.35$, as in both cases the turning surface occupies radii $r
  > 0.35 R$.  Differences for $\Omega_p \lesssim 1.4 \Omega_{\rm ref}$
  are essentially through the strong attractor at $\Omega_p \sim 0.8
  \Omega_{\rm ref}$, which is related to a turning surface located at
  $r < 0.35 R$.

  Although the above results have been obtained with a fixed profile
  of shellular rotation, the general conclusion that can be drawn is
  that, just like for solid-body rotation, a differentially rotating
  spherical shell may harbour both short-period attractors {and}
  quasi-periodic orbits of characteristics. Their occurrence depends
  on the rotation profile and the sphere's aspect ratio. An example of
  $m=0$ D mode with a weak attractor, obtained for $E=10^{-8}$,
  $\sigma=0.5$ and $\eta=0.35$, is shown in the bottom-left panel of
  Fig.~\ref{quasiper}, and it is clear that the kinetic energy for
  this mode is more homogeneously distributed than in the modes
  illustrated in previous figures.
  
  While the calculation of Lyapunov coefficients unambiguously shows
  the existence of quasi-periodic orbits of characteristics, it is
  difficult to use it to prove or disprove the existence of regular
  modes for which $\Lambda$ should strictly vanish \citep{Dintrans99,
    RGV01}. We have looked for regular modes with differential
  rotation, though. It was suggested by \cite{RGV01} that toroidal
  modes, with Doppler-shifted eigenfrequencies of the form
  $\tilde\Omega_p = 2\Omega / (m+1)$ ($m$ is the azimuthal wavenumber)
  are the only regular modes in a rigidly-rotating spherical shell. We
  have followed the toroidal mode with $m=3$ by slowly increasing the
  differential rotation rate (for shellular rotation).  As shown in
  the bottom-right panel of Fig.~\ref{quasiper}, even for values of
  $\sigma$ as small as a few percent, the mode's regularity fades
  away. Although most of the kinetic energy remains contained in a
  smooth structure reminiscent to that of the toroidal mode, ray
  structures develop, in particular from the inner critical
  latitude. Therefore, inertial oscillations with differential
  rotation seem to be all singular in the limit of vanishing Ekman
  numbers.  A similar result was obtained for gravito-inertial modes
  \citep{Fried82, Dintrans99}, which also satisfy a PDE of mixed type,
  analogous to Eq.~(\ref{poincare2}).  }

\subsection{Non-axisymmetric eigenmodes: corotation resonances}
\label{sec:num_nonaxi}
We briefly examine in this section the behavior of viscous,
non-axisymmetric inertial waves when their frequency in the rotating
frame ($\tilde\Omega_p$) vanishes inside the shell.  Such corotation
resonances will be referred to as critical cylinders for cylindrical
rotation, and as critical layers for shellular rotation.  We present
results of linear numerical calculations, which we compare to the
analysis of Section~\ref{sec:inv}. Cylindrical and shellular rotation
profiles are considered in Sections~\ref{sec:num_nonaxi_cyl}
and~\ref{sec:num_nonaxi_shell}, respectively.

\subsubsection{Cylindrical rotation}
\label{sec:num_nonaxi_cyl}
\begin{figure}
  \centering
  \resizebox{\hsize}{!}
   {
    \includegraphics{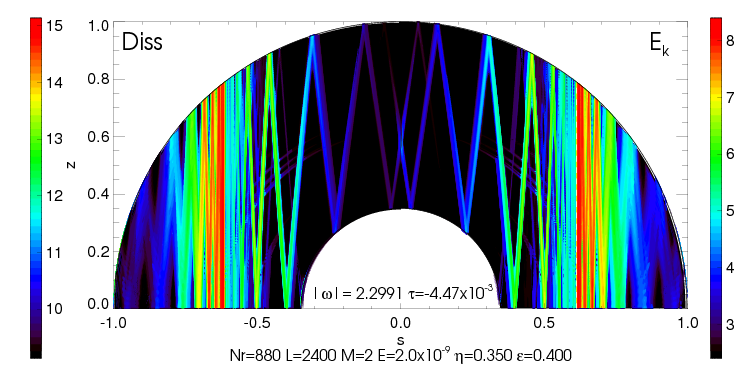}
   }
   \resizebox{\hsize}{!}
   {
    \includegraphics{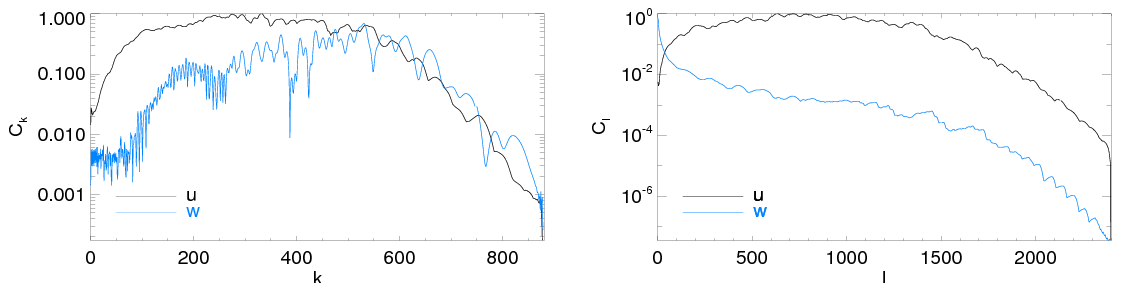}
  } 
  \caption{\label{nice5_cyl}Top: meridional cut of the viscous
    dissipation (left quadrant) and kinetic energy (right quadrant)
    for a D mode with $m=2$, eigenfrequency $\Omega_p \approx -2.3
    \Omega_{\rm ref}$, $E=2\times 10^{-9}$ and cylindrical rotation
    profile $\Omega(s) = 1 + 0.4 s^2$. A corotation resonance
    (critical cylinder) is located at $s \approx 0.6$.  Bottom:
    spectral content of the radial ($u$) and orthoradial ($w$)
    components of the velocity field for this mode. Chebyshev and
    spherical harmonic coefficients are shown in the left and right
    panels, respectively.}
\end{figure}
An example of $m=2$ D eigenmode with a critical cylinder is displayed
in Fig.~\ref{nice5_cyl}. This mode is obtained for $\varepsilon =
0.4$, $E=2\times 10^{-9}$, and the very high spectral resolution $N_r
= 880 \times L = 2400$.  Its eigenfrequency $\Omega_p \approx -2.3
\Omega_{\rm ref}$ (a negative $\Omega_p$ is required for an $m>0$
eigenmode to have a corotation resonance within the shell). The top
panel of Fig.~\ref{nice5_cyl} shows a meridional cut of the viscous
dissipation (left quadrant) and of the kinetic energy (right quadrant)
for this mode. The mode's spectral content is also displayed in the
bottom panel of Fig.~\ref{nice5_cyl} {\color{black} to highlight the
  good convergence in spectral resolution obtained for this
  mode}. According to Eq.~(\ref{scrit}), a critical cylinder is
located at $s_{\rm crit} \approx 0.61 R$.  Near this location, the
mode's kinetic energy and viscous dissipation are discontinuous. Upon
approaching corotation from $s > s_{\rm crit}$, the mode's pattern in
a meridional {\color{black}quarter-plane} becomes more and more
parallel to the rotation axis, and according to
Eq.~(\ref{cyl_vgroup}), the component of the group velocity
perpendicular to the critical cylinder should eventually cancel out at
corotation, formally preventing the mode from crossing the critical
cylinder. Our simulation indicates that, in a moderately viscous
fluid, waves get dramatically absorbed upon crossing corotation, but
keep on propagating at $s < s_{\rm crit}$ with a much reduced
amplitude.

\begin{figure}
  \centering
  \resizebox{\hsize}{!}
   {
    \includegraphics{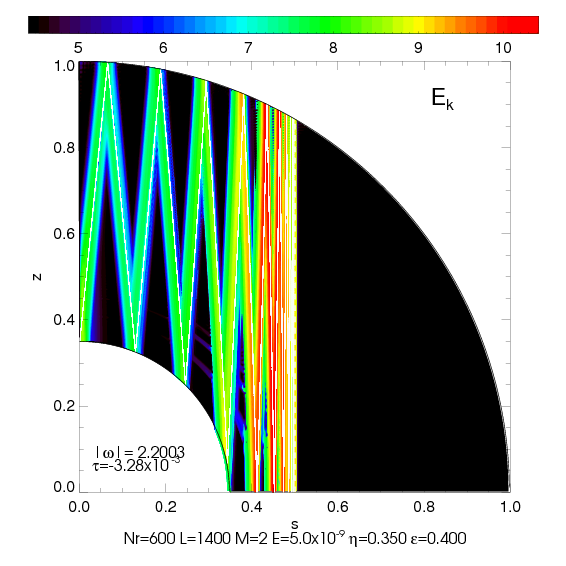}
    \includegraphics{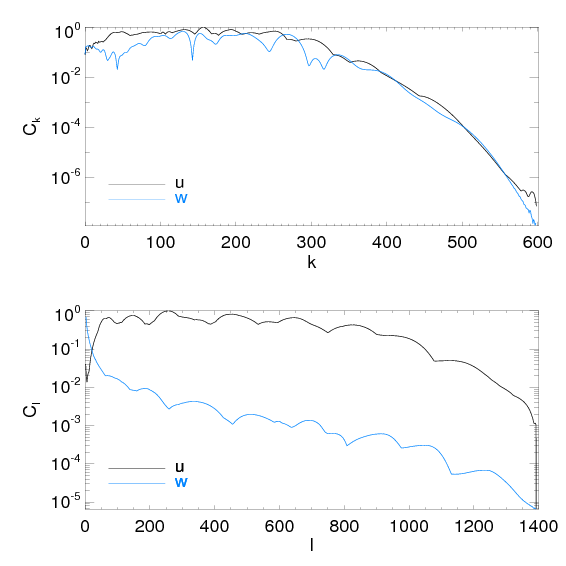}
  }
  \caption{\label{nice4_cyl}Left: meridional cut of the kinetic energy
    for a D mode with $m=2$, eigenfrequency $\Omega_p \approx -2.20
    \Omega_{\rm ref}$, $E=5\times 10^{-9}$, and $\varepsilon = 0.4$. A
    corotation resonance is located at $s \approx 0.5$. A few paths of
    characteristics are overplotted by white curves. Right: spectral
    content of the radial ($u$) and orthoradial ($w$) components of
    the velocity field for this mode.  Chebyshev and spherical
    harmonic coefficients are shown in the top and bottom sub-panels,
    respectively.}
\end{figure}

We point out that the fraction of the mode's amplitude that is
effectively transmitted across corotation depends on the mode of
interest. A similar $m=2$ mode with eigenfrequency and Ekman number
slightly increased to $\Omega_p \approx -2.20 \Omega_{\rm ref}$ and
$E=5\times 10^{-9}$, respectively, shows no significant transmission
in Fig.~\ref{nice4_cyl}, as if the incoming wave was fully absorbed at
corotation. The contour plot in the left panel of this figure gives an
upper limit for the kinetic energy transmission factor $\sim 10^{-6}$.
{\color{black} Note that in the example shown in Fig.~\ref{nice4_cyl},
  linear inertial waves propagate from the inside outwards, whereas
  Fig.~\ref{nice5_cyl} shows an opposite direction of propagation
  (from the outside inwards). It is likely due to a mode selection,
  with different modes existing on each side of the corotation
  radius.}

\subsubsection{Shellular rotation}
\label{sec:num_nonaxi_shell}
\begin{figure}
  \centering
   \resizebox{\hsize}{!}{
        \includegraphics{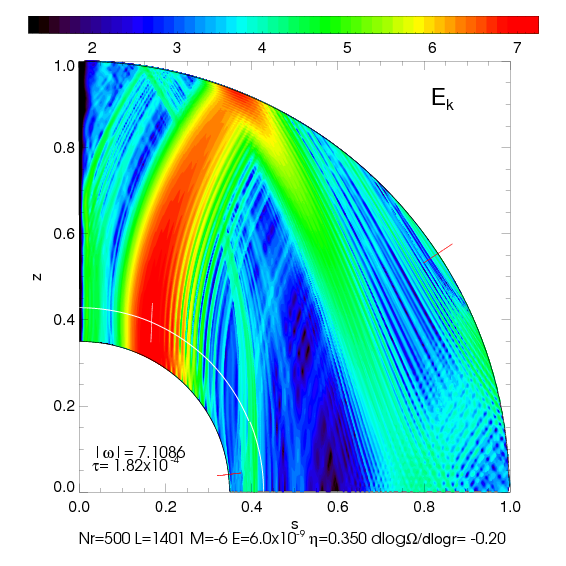}
        \includegraphics{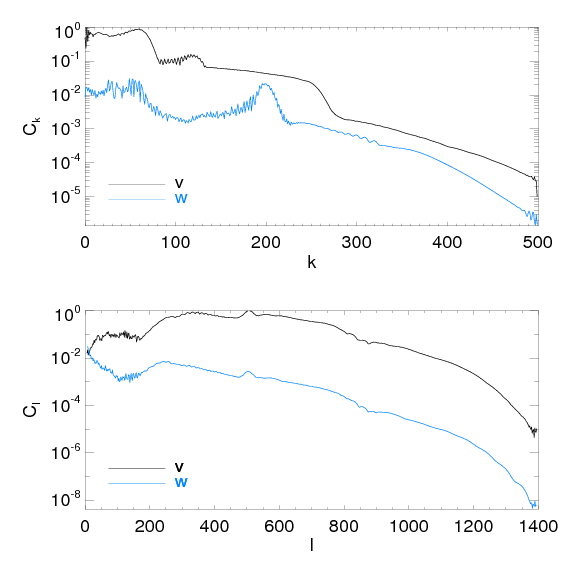}
    }    
    \caption{\label{m1}Left: meridional cut of the kinetic energy for
      a growing eigenmode with $E=6\times 10^{-9}$, $\sigma=-0.2$,
      $\Omega_p \approx 7.11 \Omega_{\rm ref}$ and $m=-6$. The
      location of the critical layer is depicted by a solid curve,
      along which a white tick shows the local slope of
      characteristics given by Eq.~(\ref{slopecrit}). Red ticks show
      the inner and outer critical latitudes. Right: spectral content
      of the radial (the quantity $v=xu$ is displayed here) and
      orthoradial ($w$) components of the velocity field for this
      mode. Chebyshev and spherical harmonic coefficients are in the
      top and bottom sub-panels, respectively.}
\end{figure}
An example of $m=-6$ eigenmode with a critical layer is depicted in
Fig.~\ref{m1}. For this mode, $E=6\times 10^{-9}$, $\Omega(r) \propto
r^{-0.2}$, and the spectral resolution is $N_r = 500 \times L = 1401$.
This mode's eigenfrequency $\Omega_p \approx 7.11 \Omega_{\rm ref}$
(this time, a positive $\Omega_p$ is required for an $m<0$ eigenmode
to have a corotation resonance within the shell). Contours of the
kinetic energy in a meridional {\color{black}quarter-plane} are shown
together with the mode's spectral content. The critical layer is
depicted by a white solid curve in the left panel, and we see that, in
contrast to cylindrical rotation, waves cross the critical layer
without experiencing significant absorption. The local analysis
carried out for the inviscid case in Section~\ref{sec:inv_corot_shell}
shows that (linear) inertial waves may indeed cross a critical layer
with a (formally) divergent group velocity at the critical layer.
This occurs if $|k_z| \rightarrow 0$ (paths of characteristic then
being locally parallel to the rotation axis), or if ${\cal B}
\rightarrow 0$ (paths of characteristics having a finite slope at the
critical layer). The white tick displayed on top of the critical layer
in the left panel of Fig.~\ref{m1} shows the local slope of
characteristics if ${\cal B} = 0$, which is given by
Eq.~(\ref{slopecrit}). The large slope obtained in this case (due to
the small value of $|\sigma|$) prevents a clear distinction with the
case $|k_z| \rightarrow 0$. The local analysis also shows the
possibility that the group velocity formally vanishes at a critical
layer (when $|k_s| \rightarrow \infty$ locally), much like in the case
of cylindrical rotation. However, we have not found such situation in
our numerical exploration of modes featuring a critical layer.

\begin{figure}
  \centering
   \resizebox{0.5\hsize}{!}{
        \includegraphics{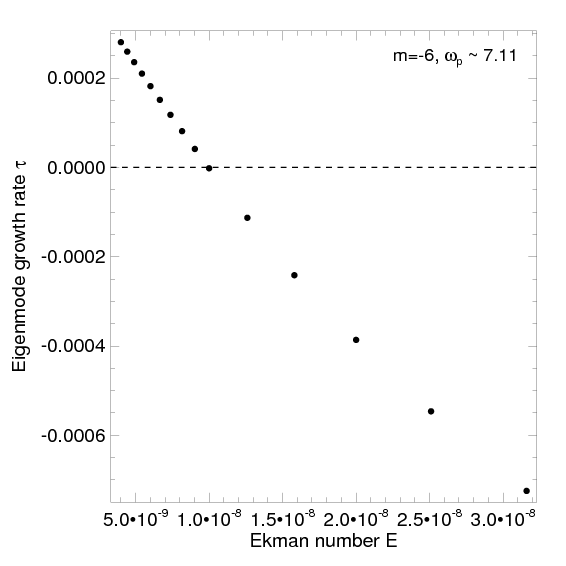}
    }
    \caption{\label{m1suivi}Growth rate for the same eigenmode as in
      Fig.~\ref{m1} ($m=-6, \sigma = -0.2, \Omega_p \approx 7.11
      \Omega_{\rm ref}$) with varying the Ekman number $E$. The mode
      becomes unstable for $E \lesssim 10^{-8}$.}
\end{figure}

The most salient feature of the mode shown in Fig.~\ref{m1} is that it
is found to be unstable: its growth rate, $\tau$, is positive. We have
examined in some more details this particular mode by varying the
Ekman number, keeping the same spectral resolution. We have varied $E$
in the range $[4\times 10^{-9}-3.2\times 10^{-8}]$. Results are shown
in Fig.~\ref{m1suivi}, where the mode's growth rate is depicted as a
function of the Ekman number. There is a transition occurring near
$E_c \approx 10^{-8}$: $\tau < 0$ (the mode is decaying) for $E >
E_c$, whereas $\tau > 0$ (the mode is growing) for $E < E_c$. The
mode's growth rate is actually well approximated by a linear function
of $E$ near this transition: $\tau = \alpha (E - E_c)$, with $\alpha <
0$.  Although not displayed here, the kinetic energy obtained in the
simulations with $E \gtrsim 10^{-8}$ is very similar to that of the
left panel in Fig.~\ref{m1}, except for the thick shear layer on each
side of corotation, which is found to progressively disappear with
increasing $E$. {\color{black} The very good convergence in spectral
  resolution obtained in these calculations (as is highlighted by the
  right panels in Fig.~\ref{m1}), and the fact we have been able to
  successfully follow this particular mode with decreasing Ekman
  number (see Fig.~\ref{m1suivi}) makes us confident that the positive
  growth rate that we obtain at small Ekman numbers is physical, and
  not a numerical artifact arising from truncature or round-off
  errors. It is tempting to interpret this result based on the local
  dispersion relation obtained at Eq.~(\ref{disp1}). It is possible
  indeed that the instability we have obtained in our simulations is
  related to the Goldreich-Schubert-Fricke instability in the limit of
  an incompressible inviscid fluid. The physical mechanism that drives
  the instability in our model requires a specific investigation that
  is postponed for future work}.

\section{Astrophysical implications and concluding remarks}
\label{sec:discu} 
The late evolution of the orbital elements and spin of short-period
extrasolar planets, or of close binary-star systems, is primarily
driven by tidal interactions. The tidal torque acting on a
gravitationally perturbed body encompasses the equilibrium torque,
which arises from the quasi-hydrostatic tidal bulge setting around the
body, and the dynamical torque, due to tidally forced internal waves
propagating inside the body. Inertial waves may propagate in
convective regions of stars and planets, where the Coriolis force is
the main restoring force. Gravito-inertial waves may propagate in
radiative zones, where buoyancy and rotation act as restoring
forces. The magnitude of the dynamical torque is directly related to
how much angular momentum is deposited by internal waves, which has a
complex dependency upon the forcing potential (amplitude and
frequency), and the body's internal properties \citep[see,
e.g.,][]{OL04, OL07, O09, RV10}. In many dynamical studies of
short-period exoplanets, the tidal torque of the central star is
parametrized by a single, dimensionless tidal quality factor $Q$,
which should be thought of, and determined as, a frequency-averaged
quantity \citep{O09}.

Prior to an investigation of tidally-forced inertial waves in
differentially rotating stars and planets, we have examined the
propagation properties of linear inertial waves in a rotating fluid
body contained in a spherical shell, taking into account the effects
of differential rotation. We have used a simple model that consists of
an incompressible homogeneous fluid, with dissipation occurring
through viscous forces. This fluid model is an intentionally
simplified model for the convective zones of stars and planets.  We
have considered two particular rotation profiles: a cylindrical
rotation profile, and a shellular rotation profile. For the sake of
simplicity, we have assumed the background flow to be in equilibrium
with differential rotation. In the case of shellular rotation, this
equilibrium would require additional forces that we have not included.
In this respect, the relaxation of the incompressible assumption, or
the generalization to stratified fluid bodies, will certainly be an
improvement on our model.

Upon linearization of the governing fluid equations, we have revisited
the propagation properties of paths of characteristics with
differential rotation, in the inviscid limit. Paths of characteristics
are shown to obey a second-order PDE of mixed type. Two kinds of
inertial modes are thus found to exist with differential rotation: D
modes, for which the hyperbolic domain covers the entire shell, and DT
modes, for which the hyperbolic domain covers part of the shell. We
have analyzed the occurrence of axisymmetric ($m=0$) and
non-axisymmetric D and DT eigenmodes for various differential rotation
rates and eigenfrequencies. The presence of turning surfaces for DT
modes extends the range of frequencies at which inertial modes of
oscillation may {\color{black}exist} compared to solid-body
rotation. We have also discussed the presence of critical latitudes
and corotation resonances with differential rotation.

The analytic results based on the dynamics of characteristics are then
compared to the results of numerical
simulations of linear inertial waves propagating in a viscous fluid 
{\color{black}contained in a spherical shell}.
High-resolution calculations, based on a spectral method, have been
performed to properly describe the structure of the thin shear layers
arising in the physically interesting regime of small Ekman numbers
(small viscosities). The Ekman numbers used in our simulations
typically range from a few $10^{-9}$ to $10^{-6}$, values that are,
however, much larger than what may be expected in stars and planets
(Ekman numbers between $10^{-15}$ and $10^{-10}$). The modes that we
have primarily focused on are singular modes with a shear layer
converging towards a short-period wave attractor. As in the solid-body
rotation case, the pattern of these shear layers is well reproduced by
the dynamics of characteristics under the short-wavelength
approximation. As an interesting side result, we have shown that
reflexions on a turning surface, and on the body's outer radius, may
form {\color{black}singular modes with short-period} wave attractors in 
the whole sphere (that is, in the absence of an inner core). 
The damping rate of singular modes with differential
rotation is also assessed in the limit of small Ekman numbers. Our
results of calculations indicate that an $E^{1/3}$ scaling is to be
expected for most D and DT eigenmodes, much similar to the case of
solid-body rotation. Still, the case of a DT mode with damping rate
approximately independent of the Ekman number is uncovered. 
{\color{black}In addition, we have carried out a qualitative analysis 
of the occurrence of modes with quasi-periodic orbits of characteristics 
for shellular rotation.
}

We have also briefly examined the behavior of linear inertial waves
with a corotation resonance within the shell (where the wave's
frequency vanishes in the frame rotating with the fluid). For
cylindrical rotation, our simulations show that inertial waves may
cross a corotation resonance, but experience a very large absorption
at this location, much like inertial waves propagating in
differentially rotating astrophysical discs \citep{LA09}. For
shellular rotation, however, we have found that waves may propagate
across a corotation resonance (critical layer) without visible
absorption. Such modes are found to be unstable for small enough
viscosities. Clearly, these findings appeal for a confirmation with an
improved fluid model, where shellular rotation is balanced for by
appropriate body-forces, and where non-linear effects are taken into
account.

Many simplifying assumptions in the fluid model we have considered
deserve further investigation to make it more applicable to the fluid
flows inside stars or planets. The incompressible and homogeneous
assumptions are obvious examples, and naturally a full model including
appropriate modeling for the convective and radiative zones would be
desirable. Since inertial waves propagate in convective regions of
stars and planets, it would be interesting to study the impact of
convective turbulence on the propagation and dissipation properties of
inertial waves (the formation of large-scale attractors for
instance). The nature and quality of wave reflections at the shell's
boundaries should be analyzed, as it is still unclear for example how
inertial waves reflect on the upper atmosphere of stars and planets.
\\
\par We thank Henrik Latter, Richard Nelson and Gordon Ogilvie for
very helpful discussions. MR would like to specially thank Susan
Friedlander who suggested this work during a visit in Toulouse.  Many
years after, these discussions have borne fruit!  We are also grateful
to the three anonymous referees for their constructive
reports. Numerical simulations were carried out on the CalMip machine
of the Centre Interuniversitaire de Calcul de Toulouse, and on the
Darwin Supercomputer of the University of Cambridge High Performance
Computing Service.  CB acknowledges support from a Herchel Smith
Postdoctoral Fellowship of the University of Cambridge.

{\color{black}
\appendix
\section{}\label{appA}
This appendix derives the partial differential equation satisfied by
the perturbed pressure, $p$, in the inviscid limit. It also derives
the wave dispersion relation, the phase and the group velocities under
the short-wavelength approximation.  In cylindrical coordinates
$(s,\varphi,z)$, Eq.~(\ref{eq:mom}) becomes
\begin{equation}
i\tilde{\Omega}_p \vect{u} + 2\Omega
  \vect{e_z} \times \vect{u} + s (\vect{u} \cdot
  \vect{\nabla}\Omega) \vect{e_{\varphi}} = \vect{\nabla}p,
  \label{eqmom_app}
\end{equation}
with $\Omega$ the background angular frequency and $\tilde\Omega_p$
the Doppler-shifted frequency.  The cylindrical components of the
linearized velocity vector, $(u_s, u_{\varphi}, u_z)$, can be
expressed as a function of $p$ and of its partial derivatives:
\begin{equation}
  u_s (s,z) = \frac{1}{i\tilde{\Omega}_p (A_s - \tilde{\Omega}_p^2)} 
  \times
  \left[
    \tilde{\Omega}_p^2\frac{\partial p}{\partial s} 
    + A_z \frac{\partial p}{\partial z} 
    + 2\Omega\tilde{\Omega}_p \frac{mp}{s}
  \right],
\label{eq_us_app}
\end{equation}
\begin{equation}
  u_{\varphi} (s,z) = \frac{1}{2\Omega (A_s - \tilde{\Omega}_p^2)} 
  \times
  \left[
    A_s \frac{\partial p}{\partial s}
    + A_z \frac{\partial p}{\partial z} 
    + 2\Omega\tilde{\Omega}_p \frac{mp}{s}
  \right],
\end{equation}
\begin{equation}
  u_{z} (s,z) = \frac{i}{\tilde{\Omega}_p} \frac{\partial p}{\partial z},
\label{eq_uz_app}
\end{equation}
and 
where 
\begin{equation} A_s = \frac{2\Omega}{s}\frac{\partial}{\partial s}
  \left( s^2 \Omega \right) \;\;\;\;{\rm and} \;\;\;\; A_z =
  \frac{2\Omega}{s}\frac{\partial}{\partial z} \left( s^2 \Omega
  \right).
\label{eq_cylkappas_app} 
\end{equation}
For solid-body rotation or cylindrical rotation, $A_z = 0$ and $A_s =
4\Omega^2$.  Note the special case where $\tilde{\Omega}_p = 0$ for
which a corotation resonance exists in the shell, and that where $A_s
= \tilde{\Omega}_p^2$, which is analogous to Lindblad resonances in
compressible fluids with cylindrical rotation such as astrophysical discs.

Taking the divergence of Eq.~(\ref{eqmom_app}) and using
Eqs.~(\ref{eq_us_app}) to~(\ref{eq_uz_app}), we find, after some
algebra, that $p$ satisfies:
\begin{eqnarray}
  \lefteqn{
    -\tilde{\Omega}_p^2 \frac{\partial^2 p}{\partial s^2}
    + (A_s - \tilde{\Omega}_p^2) \frac{\partial^2 p}{\partial z^2}
    - A_z \frac{\partial^2 p}{\partial s \partial z}
    =
  }\label{eq_poin_app}\\
  &&   
  \;\;\;\frac{\partial p}{\partial s} \left[
    \frac{\tilde{\Omega}_p^2}{s}
    +
    \frac{\partial A_s}{\partial s}
    -
    \frac{mA_s\tilde\Omega_p}{2\Omega s}
    -
    \frac{A_s}{A_s - \tilde{\Omega}_p}\frac{\partial}{\partial s}(A_s - \tilde{\Omega}_p^2)
  \right]
  \nonumber\\
  &&
  +\frac{\partial p}{\partial z} \left[
    \frac{A_z}{s}
    +
    \frac{\partial A_z}{\partial s}
    -
    \frac{mA_z\tilde\Omega_p}{2\Omega s}\left( 1 + \frac{4\Omega^2}{\tilde{\Omega}_p^2} \right)
    -
    \frac{A_z}{A_s - \tilde{\Omega}_p^2}\frac{\partial}{\partial s}(A_s - \tilde{\Omega}_p^2)
  \right]
  \nonumber\\
  &&
  +\frac{mp}{s} \left[
    \frac{m}{s} \left( A_s - \tilde{\Omega}_p^2 - 8\Omega^2 \right)
    +
    \frac{A_s \tilde{\Omega}_p}{\Omega s} \left( 1 - \frac{4\Omega^2}{A_s} \right)
    -
    \frac{2\Omega\tilde{\Omega}_p}{A_s - \tilde{\Omega}_p^2}\frac{\partial}{\partial s}(A_s - \tilde{\Omega}_p^2)
  \right],
  \nonumber
\end{eqnarray}
which reduces to Eq.~(\ref{poincare2}) when only second-order terms
are retained (the short-wavelength approximation).

We now examine the propagation properties of local waves in a
meridional plane. Assuming $p \propto \exp(i \{ k_s s + k_z z\})$,
Eq.~(\ref{eq_poin_app}) yields the following wave dispersion relation
in the short-wavelength approximation:
\begin{equation}
	\tilde\Omega_p^2 = \frac{k_z^2}{ \|{\bf k}\|^2 } 
	\left[
	A_s - \frac{k_s}{k_z} A_z
	\right],
\label{disp_app}
\end{equation}
where $\|{\bf k}\| = \sqrt{k_s^2 + k_z^2}$. For solid-body rotation,
Eq.~(\ref{disp_app}) reduces to the well-known dispersion relation
$\tilde\Omega_p = \pm 2\Omega k_z / \|{\bf k}\|$ \cite[see,
e.g.,][]{Green69}. In the bracket term of Eq.~(\ref{disp_app}), while
$A_s > 0$ is required to satisfy Rayleigh's criterion, the sign of
$A_z k_s / k_z$ is not known {\it a priori}.  Eq.~(\ref{disp_app})
therefore indicates that an instability may occur if $A_s - A_z k_s /
k_z < 0$.  It corresponds in fact to the Goldreich-Schubert-Fricke
instability for an incompressible inviscid fluid \citep{gs67,
  fricke68}, in which case the inequality $A_s - A_z k_s / k_z < 0$ is
identical to that in equation 33 (misprinted) of \cite{gs67} (see also
\cite*{NGU12}).  Assuming the fluid is stable against the
Goldreich-Schubert-Fricke instability, we introduce
\begin{equation}
\tilde{{\cal B}} = \left[ A_s - \frac{k_s}{k_z} A_z \right]^{1/2},
\end{equation}
and Eq.~(\ref{disp_app}) translates into $\tilde\Omega_p = \pm
\tilde{{\cal B}} |k_z| / \|\vect{k}\|$. In the frame rotating with the
fluid, the phase velocity, ${\bf v_p} = \tilde\Omega_p {\bf k} /
\|{\bf k}\|^2$, satisfies
\begin{equation}
	{\bf v_p} = \frac{\pm \tilde{{\cal B}} k_z }{\|{\bf k}\|^3} \,{\bf k},
	\label{vphase_app}
\end{equation}
while the group velocity, ${\bf v_g} = \partial \tilde\Omega_p
/ \partial {\bf k}$, is given by
\begin{equation}
	\vect{v_g} = \pm \frac{k_s}{\|\vect{k}\|^3} (-k_z \vect{e_s} + k_s \vect{e_z}) \times 
	\left[ \tilde{{\cal B}} + \frac{A_z}{2} \frac{\|\vect{k}\|^2}{k_s k_z} \tilde{{\cal B}}^{-1} \right].
	\label{vgroup_app}
\end{equation}
Note that ${\bf k} \cdot {\bf v_g} = 0$: the group velocity is
perpendicular to the phase velocity for differential rotation as for
solid-body rotation.  }  

\bibliographystyle{jfm}

\end{document}